# Large Magnet Structures for Fusion Technology: Lessons from ITER

*N. Mitchell*
Gauss Fusion GmbH (formerly ITER Project)

**Abstract**
Over a period of 40 years, the ITER project has provided many examples in large magnet structures, from pre-concept to design to manufacture to delivery and assembly, from which lessons can be learned for the future. This presentation is divided into three parts.

The first, 'Making and Assembling Large (steel) Cryogenic Structures for Magnets' is based on ITER Coil Structure experience, particularly the Central Solenoid (CS) and Toroidal Field (TF) structures and particularly on large 316LN forgings weighing about 50t individually, requiring then high accuracy machining and deep (up to 0.3m) welds with controlled distortion. New materials development is often proposed. This is easy on a laboratory scale, but often not transferable to an industrial large scale. Examples are provided. The issues to be considered (and solved) are large scale quality (and repair of defects), joining, tolerances and assembly.

The second 'Designing Large Steel Support Structures for Magnets' is concerned with structural design codes & criteria, and modern analysis, especially of imperfect, but manufacturable structures. Points that are considered are typical critical issues with cyclically loaded thick components, analysis of partial penetration welds and other welds not permitted under normal American Society of Mechanical Engineers (ASME) welding procedures (including fillet welds and/or those volumetrically not inspectable by Non-Destructive Examinations (NDE)) and allowing for tolerance compensation (i.e. designing with overmetal or designing for shimming).

The third 'Avoiding or Reducing Large Steel Support Structures for Magnets' shows how it is possible to choose other routes to include structural materials in magnets, for example (as in ITER TF coils) by using plates to contain the conductor, or by using a thick conductor jacket to at least share the loads with massive steel cases.

## 1 Introduction

In a magnetic confinement based Fusion Power Plant (FPP), structural materials (mostly steel) will make up about 65% of the magnet cost (actually far more than the superconductor material). As the size of fusion devices increases and they intend to reach nuclear confinement conditions, the magnets become much larger and work at higher field. Although the superconductor cost is always a very 'visible' (and easy to estimate) cost item, the magnetic forces require more structures, with more accuracy and with more gaps for maintenance. Superconductors have seen much development (and publicity) in the last 20 years with the advent of High Temperature Superconductors (HTS) which improve the operation range (field but more importantly temperature) but of course then add to the structural issues. Less visibly, magnet structures (particularly in ITER) and materials have been the beneficiaries of much steady improvement in steel forming and welding technology, although this is not very visible.

Austenitic stainless steel manufacturing technology has moved on substantially over the period of ITER design, leading to big improvements in the supply chains (although not always in time for ITER manufacturing). If ITER were to be re-designed today, it is likely that significantly different decisions would be made about the basic materials but also in the manufacturing design, even if the ITER device itself were to remain unchanged.

## 2 ITER Magnet Design

Since this paper is based on the experiences from ITER, this section includes a design explanation and performance requirements. Fig. 1 gives an overview. The TF coils (in their cases) are orange, the CS is inside the TF coil toroidal arrangement and the Poloidal Field (PF) coils (in purple) go outside the TF. Publications on the ITER magnets are available at [1] and [2].

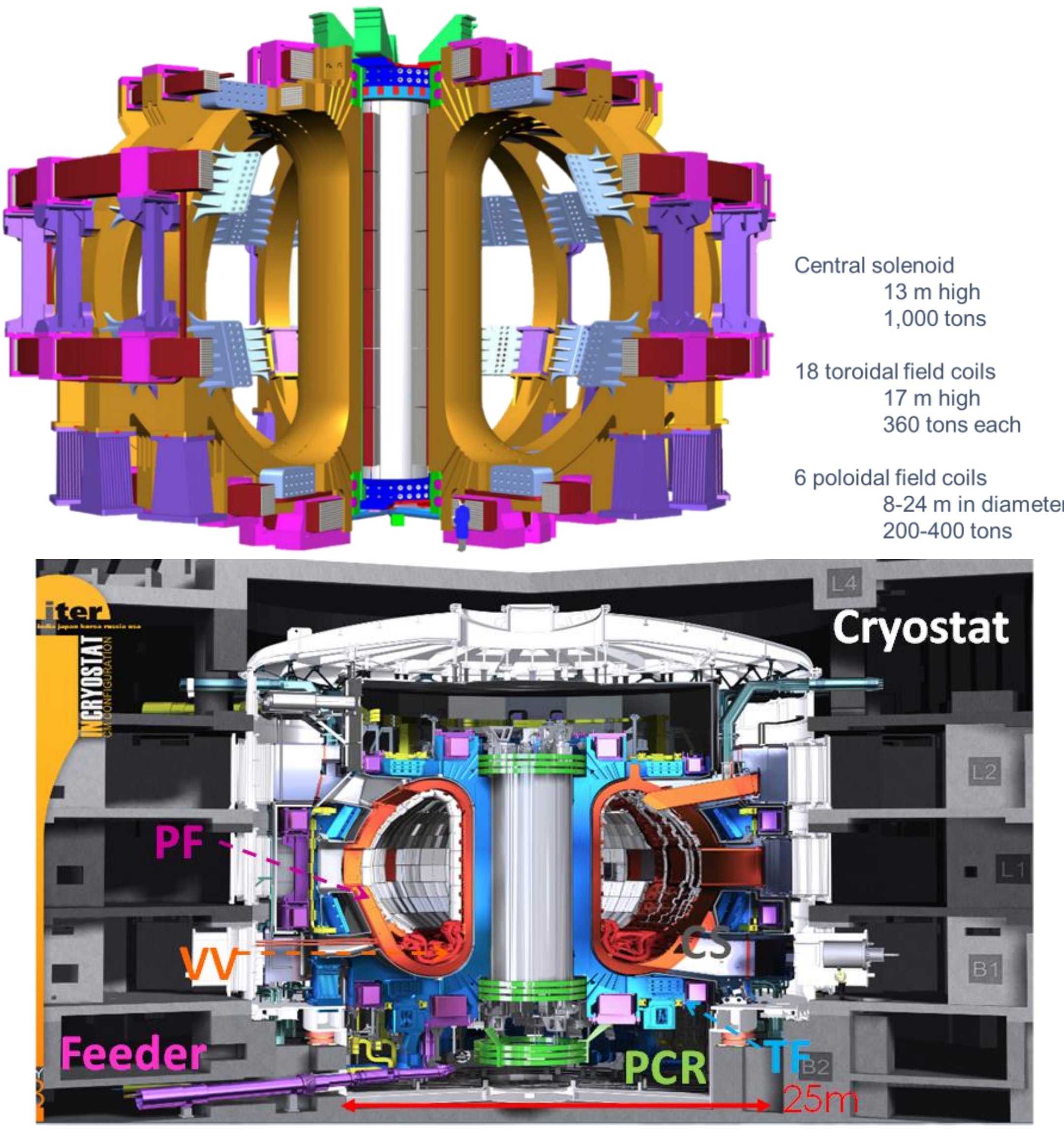

**Fig. 1:** Two overviews of the ITER Magnet System

## 2.1 Coils, Supports and Structures

There are 18 TF coils, 'D' shaped and consist of a Wing Pack (WP) enclosed in a thick steel case, and made up of 4 welded sub-assemblies. The coils are wedged together along the inner straight leg, as shown in Fig. 2.

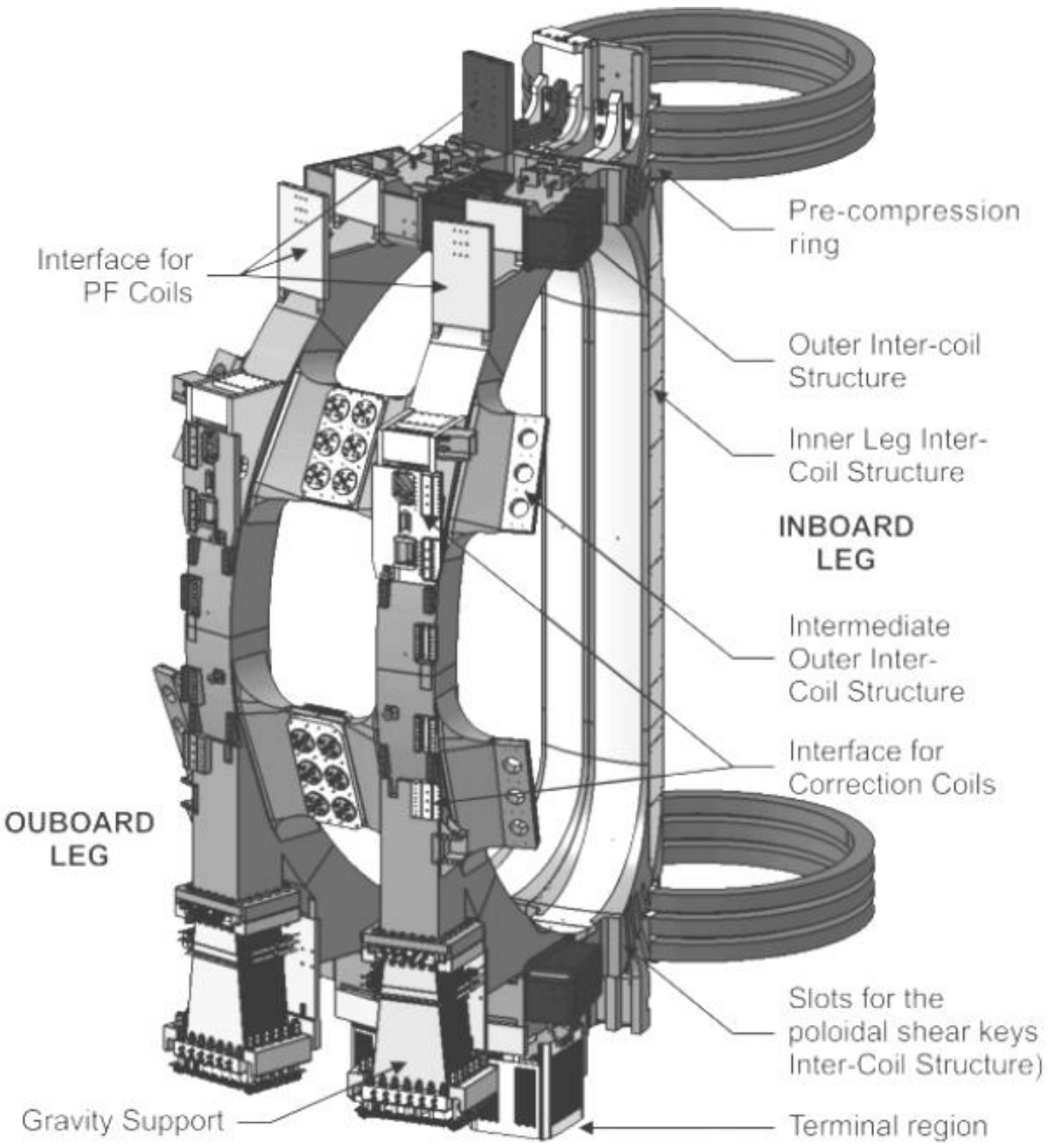

**Fig. 2:** The TF Coil Subsystem

The case is welded around the winding pack of each TF coil and is manufactured in 4 main subsections, AU, AP, BU and BP, as shown in Fig. 3.

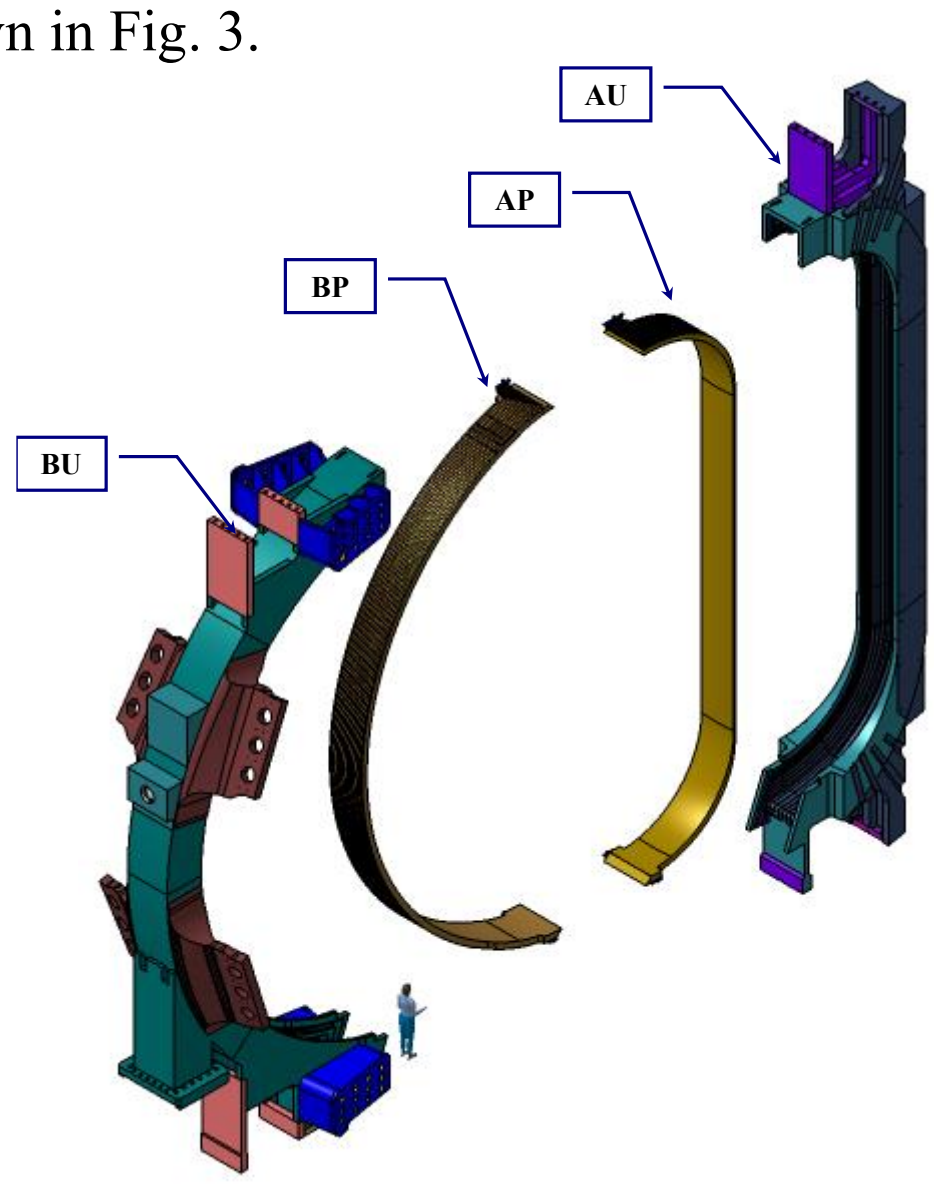

**Fig. 3:** The 4 Sub-Assemblies of the TF Coil Case

The WP, Fig. 4, is a bonded structure of 7 Double Pancakes (DP) each made up of a steel radial plate (RP), which houses the reacted Cable In Conduit Conductor (CICC), with an outer ground insulation of 7 mm thickness. The coil terminals protrude from the TF coil case at its lower curved part with the 6 DP joints (i.e. the joints linking adjacent DPs) and the helium feeder manifolds, Fig. 2.

The case U-section sub-assemblies BU and AU, Fig. 3, have numerous integral structural attachments such as the Intercoil structures, the pre-compression flanges, the TFC support leg, and attachments for the Central Solenoid (CS) coils, Poloidal Field (PF) coils as labelled in Fig. 2, as well as other stubs and holes for the Correction Coils (CC), vacuum vessel thermal shield (VVTS) supports and machine assembly tooling (not labelled in Fig. 2).

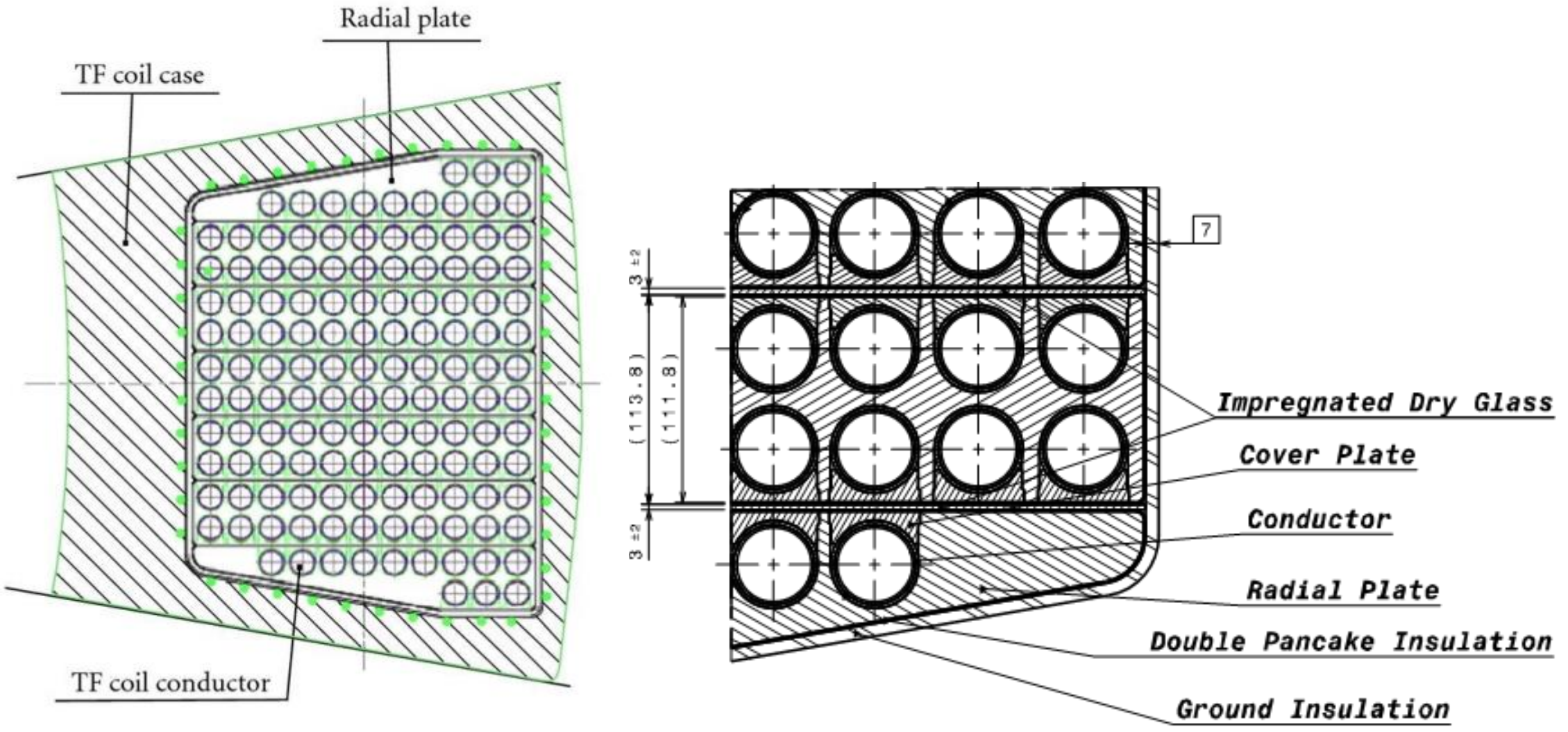

**Fig. 4:** TF Coil Cross-Section.

The CS is a stack of 6 independent circular coils, Fig. 5.

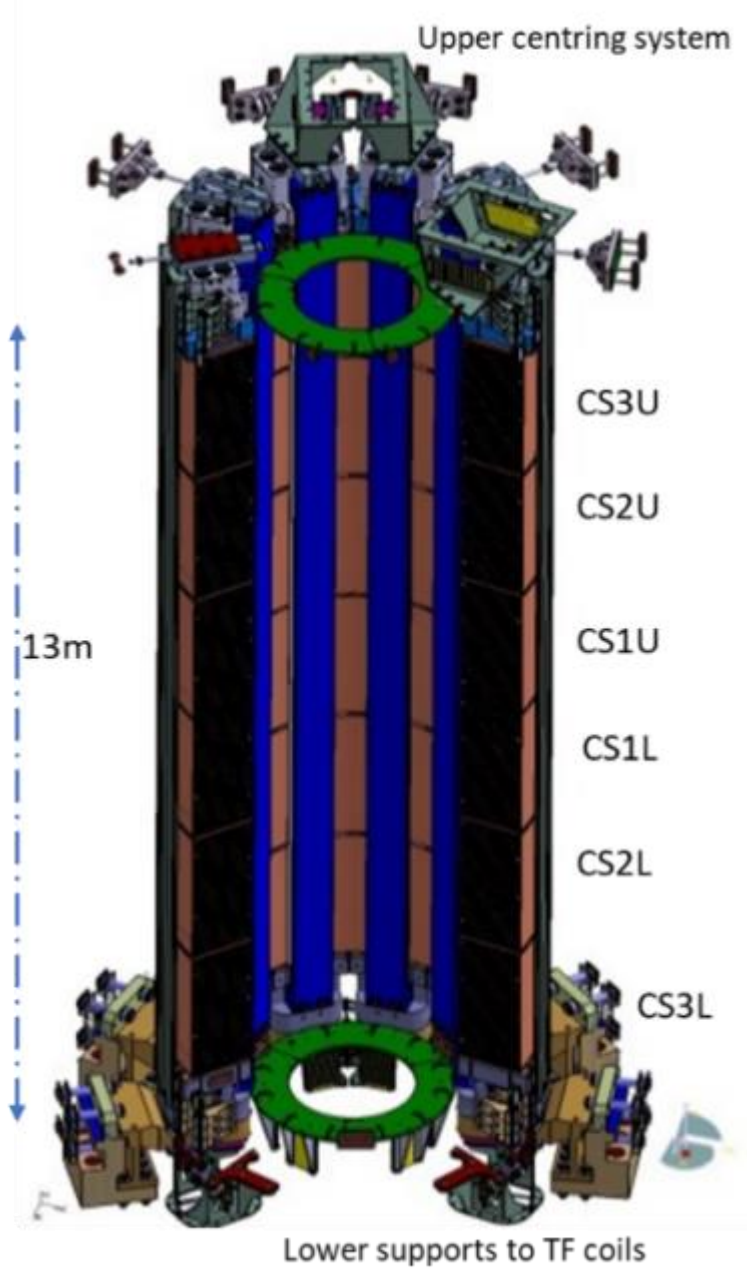

**Fig. 5:** Stack of 6 CS coils and longitudinal pre-compression flanges (9 inside, 18 outside) that link to flanges on the top and bottom that are used to put vertical pre-loading on the coil stack.

In operation, the centring force on each TF coil is reacted by toroidal hoop pressure in the central "arch" formed by the straight inboard legs of the 18 coils. The TF coil case inboard legs are wedged over their full radial thickness and about half of the centring force is reacted in toroidal direction through the WP, while the other half is reacted by the case. The front part or "nose" of each TF coil case is thicker to take the case part of this load.

When the TF coils are energized, the WP also pushes outward all round the case and tends to shrink slightly in the toroidal and radial directions due to the Poisson effect and magnetic compression. A small gap between WP and TF coil case opens on the inside (facing the plasma). This gap may locally close due to torsion of the coil under out-of-plane forces when the CS and PFC are pulsed. In the inboard leg, the wedging pressure (Section 4) forces the winding pack against the case and tends to suppress the side gaps. On the outboard leg, there is a small amount of slip between WP and TF coil case through the plasma pulse as the out-of-plane forces change.

In the inboard curved regions above and below the straight leg, the case thickness increases rapidly to provide adequate out-of-plane support. The peak stresses occur at the point where the straight leg ends and they are sensitive to the minimum wall thickness of the case at this point (this determines its torsional rigidity which is required to support the loads as the wedging stops).

### 2.2 Intercoil Structures including Precompression Rings

The ITER designs of the 1990s included elaborate methods to support the overturning moments (see Section 4) on the TF coils, partly because of the high plasma current (the moment is proportional to the current) and also because of the use of a bucked design where torsion of the CS had to be limited. This extended to full keying of the TF coils to each other in the inner leg region. The difficulty of manufacturing this to the required tolerances and then assembling it was one of the insurmountable problems that led to the wedged design to be chosen instead. By the 2001 FDR [3] these links between the individual TF coils had relaxed to 4 sets of support systems, the Inner Leg Intercoil Structure (ILIS), the Inner Support Structure keys (IIS), the OIS (outer intercoil structure) and the IOIS (intermediate OIS), plus the Pre-Compression Rings (PCRs), Fig. 2.

In the wedged design the TF coil inner legs touch and support the overturning moment by friction. This was addressed rather simplistically until 2017 without a proper design of the interface surfaces. The IIS, OIS and IOIS all went through a detailed design phase in 2010-2012. The changes were substantial, and largely addressed the issue of absorbing tolerance between the coils (match machined shells for the IIS and off set inserts and expansion bolts/pins for the OIS). The IOIS had more significant performance problems (load distribution and stress concentrations) and was completely changed, twice, from the 2001 Final Design Report (FDR) [3] design before the present system of large pins, match machined inserts and plates was developed.

The ILIS had to combine the functions of a shim (to adjust the gaps between the TF coils to achieve the required average value of 2mm for each gap (see the section below on tolerances) and to remain fixed to the coil surfaces when they are de-energised (with the risk of falling down the gap and creating a severe non uniformity in the wedging stresses), and also to provide a low voltage insulation against eddy currents. Finally, a system of variable thickness steel plates on one side and G11 high strength boards on the other, both fixed with washers, was adopted in 2018. Attaching these plates was eventually performed in a separate facility on site, partly because the design was finalised too late to be included in the work at the suppliers and partly to leave flexibility in case the ILIS needed to be customised during assembly. The original intent was that the coils could be shimmed as required (i.e.

as determined by the on going build up in the cryostat), at the last minute. This has not been possible due to schedule, order of assembly, access and orientation and all the TF coils have been prepared in the same way. It is fortunate that it seems from the manufacturing experience and the initial assembly of the first sector, that the TF manufacturing and adjustment capability during assembly is sufficiently good to avoid the need for variable ILIS shimming and the early pre-preparation will be acceptable

The ITER Pre-Compression Ring system, Fig. 2, provides the radial preload required to maintain structural integrity and coil-to-coil contact in the Toroidal Field magnet assembly. Each of the six 5.6 m fiberglass-reinforced composite rings operates at 4.2 K under sustained load for 20 years. They are formed from glass fibre impregnated into a resin matrix. The Pre-Compression Rings (PCRs) were an interesting development within ITER and the European Domestic Agency that took close to 20 years to be successfully completed, requiring 3 iterations on the way, from winding of a uniaxial glass fibre bundle to placing glass-resin tows to winding a pultruded resin-glass band. They provided many lessons regarding the development of composites as large structural load bearing components. The full history has not been published but is available in ITER internal documents [4] and Fusion for Energy (F4E) website [5]. Test results are reported in Ref. [6].

## 2.3 Distributed Structures

### *2.3.1 TF Coil Plates*

As shown in Fig. 4, each radial plate measures roughly 13.8 meters long, 8.7 meters wide, and 112 millimeters thick, weighing up to 10 metric tons and build up by EB or Laser welding from machined subsections, Fig. 6. Both sides of a radial plate feature highly precise, round-shaped spiral grooves. Jacketed $Nb_3Sn$ superconducting conductors are bent and (after an intermediate heat treatment for Nb3Sn strands) insulated and precisely fitted inside these grooves. Once the insulated conductor is laid inside the grooves on both sides, stainless steel cover plates are laser-welded over them to completely enclose the cable. This completed module is known as a Double Pancake (DP).

Structurally, the plate protects the conductor, and the brittle superconducting (sc) strands, against the accumulated magnetic forces. It also provides a highly accurate final shape to the winding. In terms of overall magnet loads (mostly the 'bursting' force) it carries about half of the load, the case carrying the rest.

In ITER the radial plates were made to the minimum size (thickness) compatible with holding the conductor, and the rest of the support was provided by the case. In terms of cost effectiveness, the plates cost about 200euro/kg in 2024 prices and the case about 90euro/kg (in both cases the weight refers to the finished product). This is not surprising but can be misleading as the cases contain very large areas of inefficiently used material (especially BU) which reduces the unit cost but increases the total. There may be advantages in future devices in eliminating the case and distributing the material into the winding pack where it is much more efficiently used (in terms of MPa/kg).

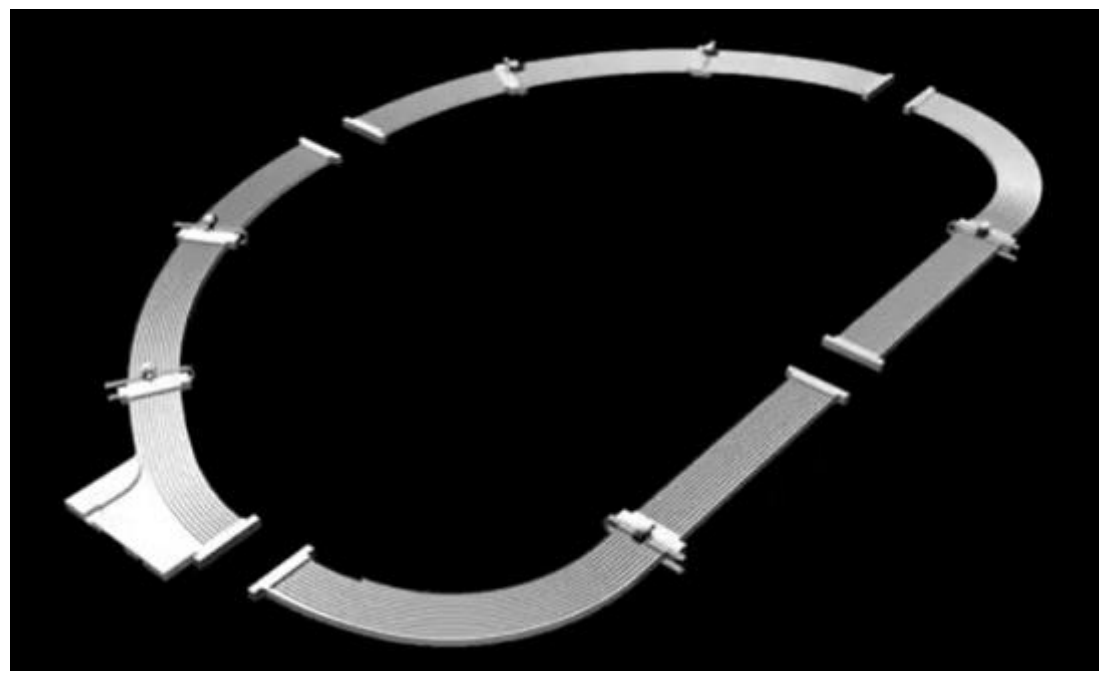

**Fig. 6:** ITER TF radial plate showing the subsections of machined rolled plates used to build it.

### *2.3.2 Conductors*

A thick conductor jacket is another route to include structural steel within a winding pack, in ITER used in the CS and PF coils. The conductor is made by pulling the completed cable into a pre-welded and oversize jacket (length up to 1400m) and then compacting the jacket onto the cable. For $Nb_3Sn$ conductor, both cable and jacket then have to go through the $Nb_3Sn$ heat treatment (about 200hrs at 650C) after final winding, which creates considerable metallurgical issues in the jacket due to carbon precipitation at grain boundaries. The jacket itself is made by extruding short sections (3-4m for heavy jackets) and then carrying out a drawing and straightening process. The final conductor is shown in Fig. 7.

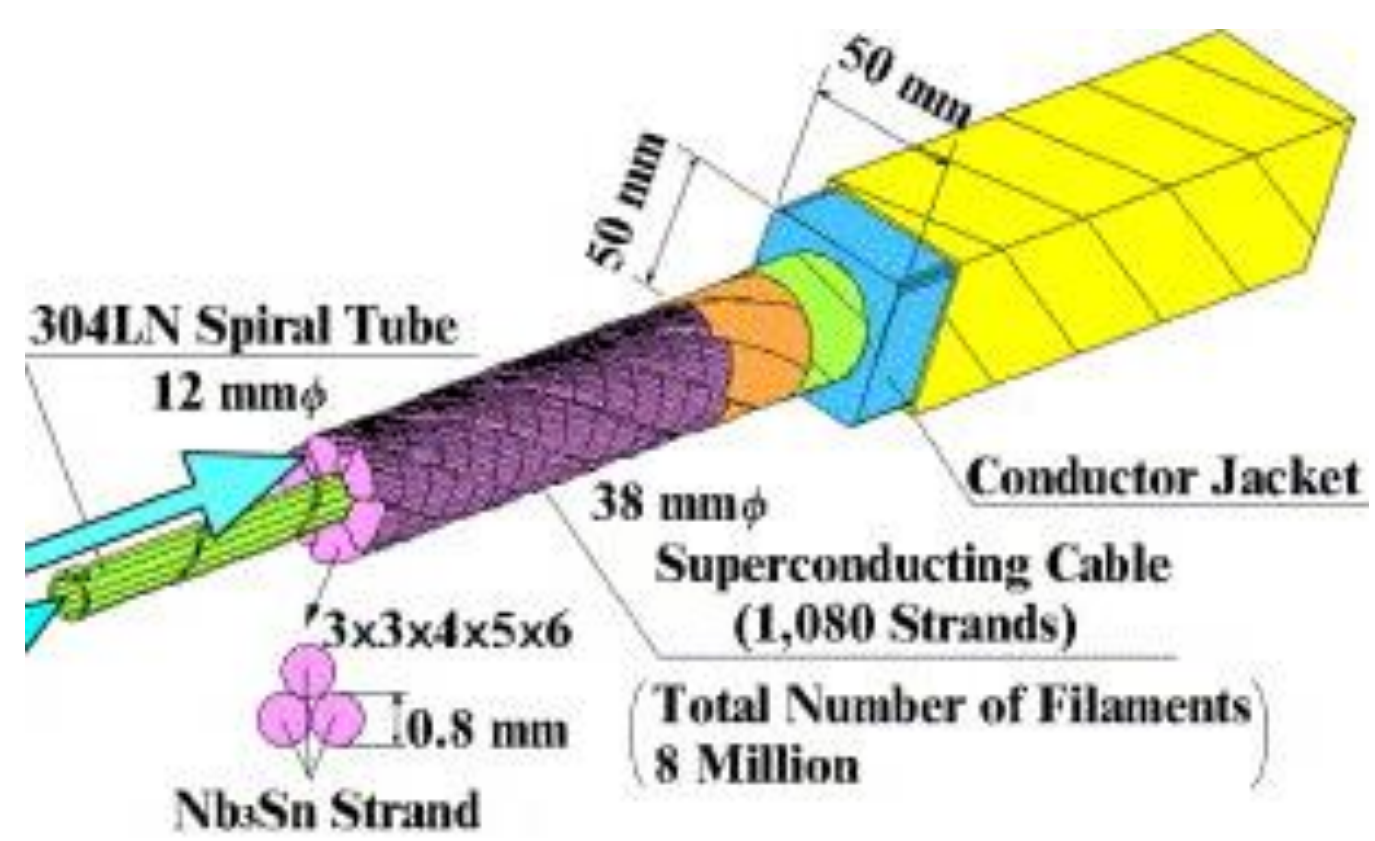

**Fig. 7**: ITER CS/PF Conductor with a square steel structural jacket.

The conductor obviously requires at least a thin jacket to contain the helium, but this can be fabricated in a circular form much more cheaply than the square jacket, in 10-12m lengths.

Considering the price of including such structural material, it is sensitively dependent on the jacket size and with an upper limit on this, but on average appears to be about 125euro/kg at 2024 prices. In general coils wound with square conductor are an effective way to include structural material but with sharp upper limits, at which point a case is needed. There has been speculation about including additional structural material in the form of a co-wound strip (which would be low cost, probably <50euro/kg) but such a coil has not been wound.

## 3 Structural Materials for Fusion Magnets

New structural materials are often suggested as a future contribution to the structural problems of FPP magnets. A look at the history of material 'innovations' for ITER over its 40 year history shows how over-simplified these proposals can be. New materials developments are often proposed, even now. It is easy on a lab scale, but often useless on a large scale as the breadth and scale (and costs) of such developments needed to bring a 'new' material to industrial availability are massively underestimated.

To start, we will look at the industrially available materials, the ITER requirements and the reasons that these materials have (or have not) been used, before looking at fusion specific efforts to improve on this.

### 3.1 Bulk Structural Materials and New Materials

With cryogenic magnet structures of thickness greater than ~20mm, the failure concern is dominated by fracture toughness and the risk of fast fracture from defects (possibly increased by fatigue). Fracture toughness is not very sensitive to temperature (and slightly decreases from room temperature with 316LN) but yield stress substantially increases. Not only are residual stresses from room temperature and above locked in, but stress relaxation at stress concentrations is reduced. So, we are concerned about cracks, either in the base material (from forging or rolling, or inclusions) or more likely in welds. This puts a high priority on weldability (i.e. a low number of defects in the first place), NDT (component geometry and material attenuation) and repair.

Then, on cost and supply chain grounds, it is difficult to move away from austenitic steels. Occasionally the Inconel family are used for very specific subcomponents (bolts and studs, keys, shims etc) where it offers improved performance over the steels. However, the difficulties in producing defect free large components and/or difficult welding have limited its larger scale use.

The present family of austenitic steels is summarised in Fig. 8 [7, 8]. Out of these, the ones that can be suitable for cryogenic applications are 304, 316LN and Nitronic 50, with various levels of low carbon (304L) and added nitrogen (which impacts in particular the welding but also machining and forging).

| Steel | C | Mn | Si | Cr | Ni | Mo | Others |
|---|---|---|---|---|---|---|---|
| 304 | 0.08 | 2 | 1 | 18–20 | 8–12 | - | - |
| 308 | 0.08 | 2 | 1 | 19–21 | 10–12 | - | - |
| 316 | 0.08 | 2 | 1 | 16–18 | 10–14 | 2–3 | - |
| 321 | 0.08 | 2 | 1 | 17–19 | 9–12 | - | 0.7 Ti |
| 347 | 0.08 | 2 | 1 | 17–19 | 9–13 | 2–3 | 1 Nb |
| 316LN | 0.02 | 1 | 0.4 | 17 | 13 | 2.2 | 0.13 N |
| 317LMN | 0.02 | 1 | 0.4 | 18.5 | 15.5 | 4.5 | 0.16 N |
| Nitronic 30 | 0.02 | 8 | 0.5 | 16 | 2.25 | - | 0.23 N |
| Nitronic 32 | 0.08 | 18 | 0.5 | 18 | - | 1 | 0.5 N; 1 Cu |
| Nitronic 40 | 0.04 | 9 | 0.5 | 20 | 6.5 | - | 0.28 N |
| Nitronic 50 | 0.04 | 5 | 0.4 | 22 | 12.5 | 2.25 | 0.3 N; 0.2 Nb |
| 254SMo | 0.01 | 0.5 | 0.4 | 20 | 18 | 6.25 | 0.2 N; 0.75 Cu |
| AL-6XN | 0.02 | 1 | 0.5 | 21 | 24.5 | 6.5 | 0.22 N |

**Fig. 8:** Chemical composition of industrial standard austenitic and superaustenitic steels, [7], wt%.

Nitronic steel was first invented in 1961 by Armco Steel. The very first grade introduced in the family of nitrogen-strengthened austenitic stainless steels was Nitronic 40. Nitronic 50 steel was first utilized and certified for cryogenic applications in 1971, shortly after its commercial release. 316LN was developed in the 1970s and became widely used in the 1980s, both in fission reactors and in cryogenic applications where it avoids the brittle transition found in 304 alloys at 4K.

The problem of weldability of austenitic steels is high sensitivity to the welding thermal cycle, grain growth, and hot cracking. Another problem is nitrogen porosity and losses in nitrogen content, especially in laser and beam welding. In general 316LN has advantages over Nitronic 50 in this respect. Nitronic 50 is significantly harder to forge because it possesses twice the flow resistance (requiring much higher press power) and demands strict temperature control to prevent hot cracking. This is likely the main reason for the prevalence of 316LN (with ultra high nitrogen) over Nitronic 50 in large cryogenic structures. Nitronic 50 offers a generally higher yield stress but lower fracture toughness which is undesirable except in some specific applications. It also requires very specific welding conditions and weld filler which limits the operational flexibility. Figure 9 shows a comparison, based on trends established in [9, 10, 11, 12]. These show trends, not specific values and 316LN in particular can show lower values: this range is typical of the ITER C2 grade (see below). The elongation to fracture is above 35% for both for the whole range, although Nitronic 50 is typically 5% lower than 316LN up to 150K.

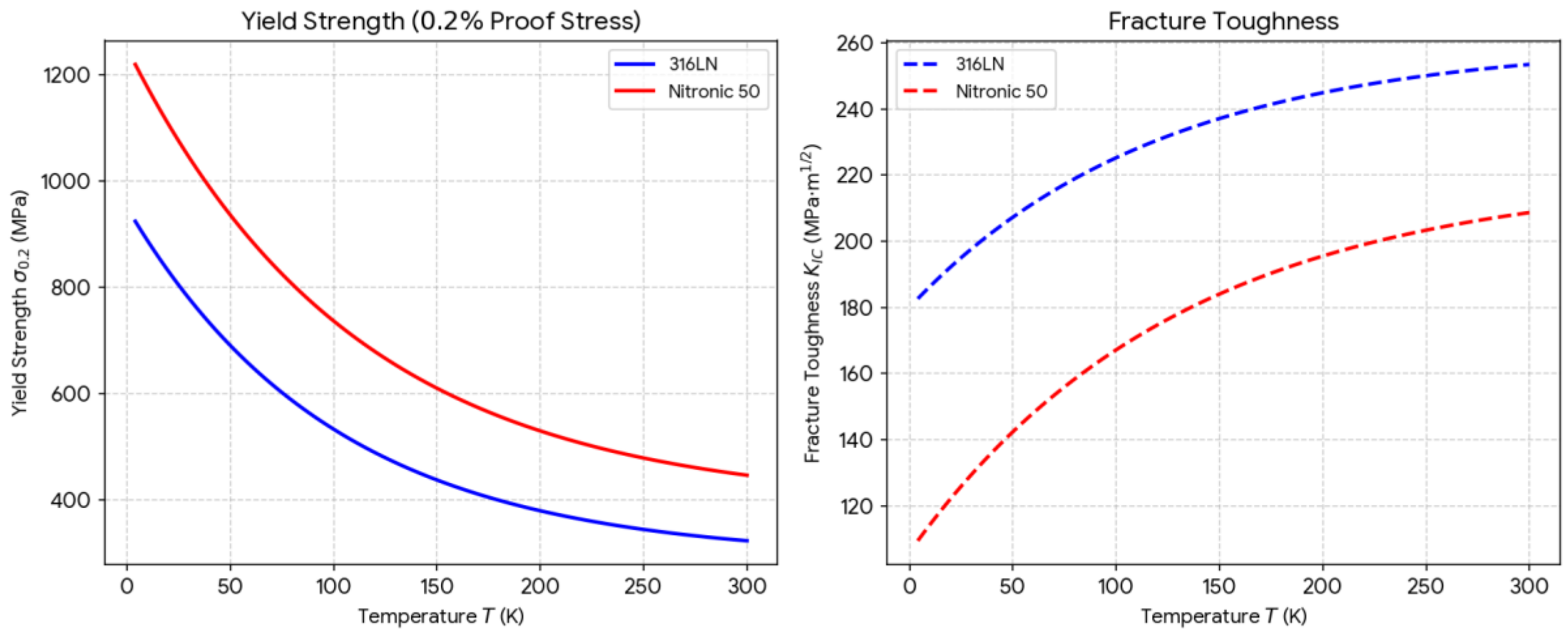

**Fig. 9:** Yield Strength and Fracture Toughness Comparison of 316LN and Nitronic 50.

In ITER Nitronic 50 steel is primarily used in the precompression tie plates of the Central Solenoid magnet system because yield stress is a priority of the design, welding is not present and the component thickness is limited.

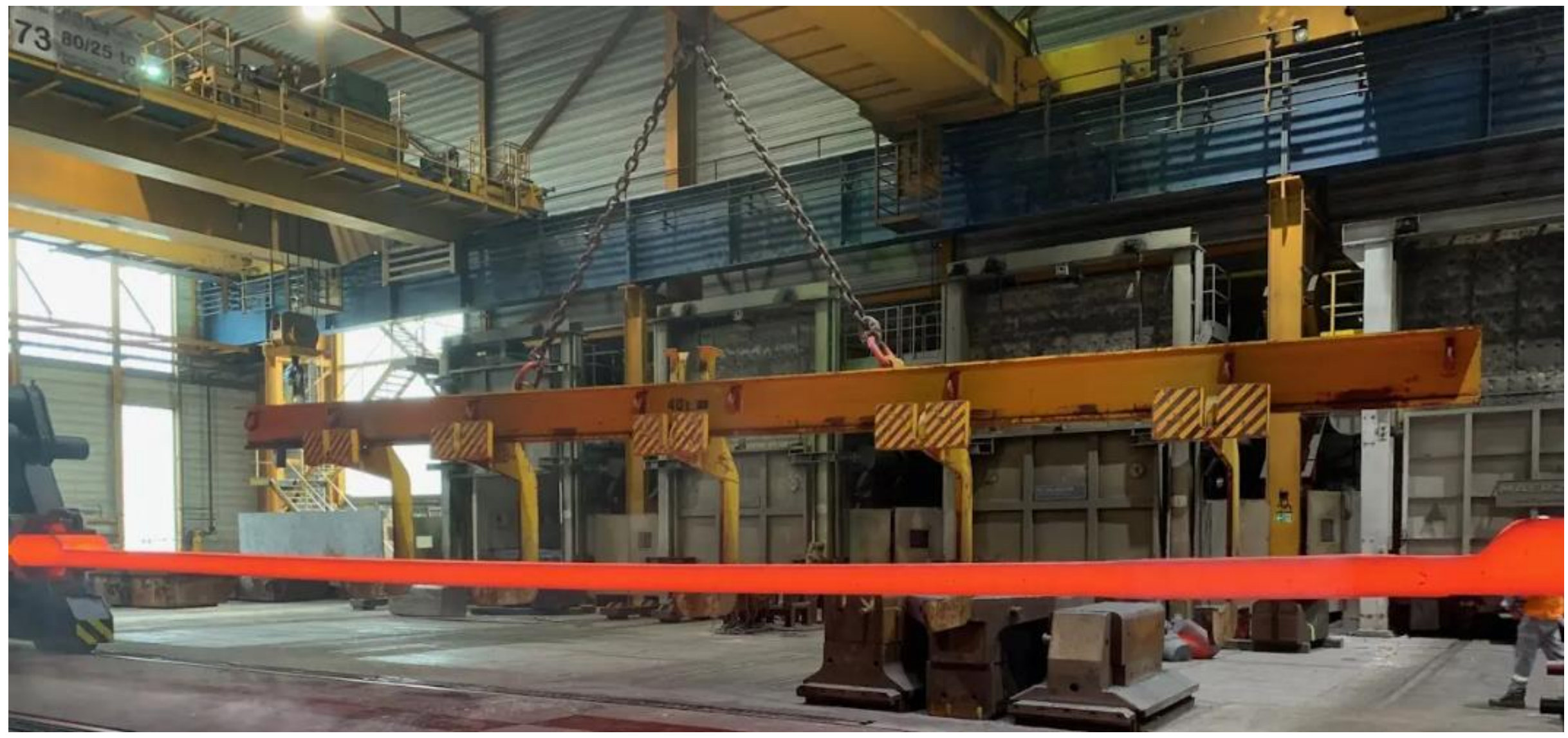

**Fig. 10:** Final forging operations of the 15 m length ITER CS tie plates. Image courtesy of Rolf Kind GmbH. [13].

Basic material research was launched in 1988 at the start of the ITER CDA, with the perception that higher structural metal properties could bring saving in overall machine cost. The program was launched in JA, EU, and RF. Success was claimed in the mid-1990s in laboratory scale research but there was universal failure on an industrial scale. Problems of production of highly composition specific alloys was underestimated (along with the willingness of steel suppliers to become involved in such a long term and small-scale development). The development also neglected issues such as welding, forging, and corrosion, so that reasonable behaviour in these areas was sacrificed to achieve a high toughness and yield.

The objectives – and results – of this development are summarised in Fig. 11. The yield stress requirement now for the highest performance stainless steel in the ITER TF structures (shown as C1) is 1000 MPa and the fracture toughness is 180 $MPam^{1/2}$, compared to a targeted development activity, in 1990, of 1200 MPa and 200 $MPam^{1/2}$, values originally thought to be attainable. The material compositions can be compared with the industrial options in Fig. 8.

By 2008, only JJ1 remained (eventually being used in the highly stressed nose region of the ITER TF coil) and steel properties were at same level as obtainable industrially in the 1980s. ITER designated 3 grades of 316LN for structural material supply, C1, C2 and C3, all lying below the 1988 trend line in Fig. 11 [2].

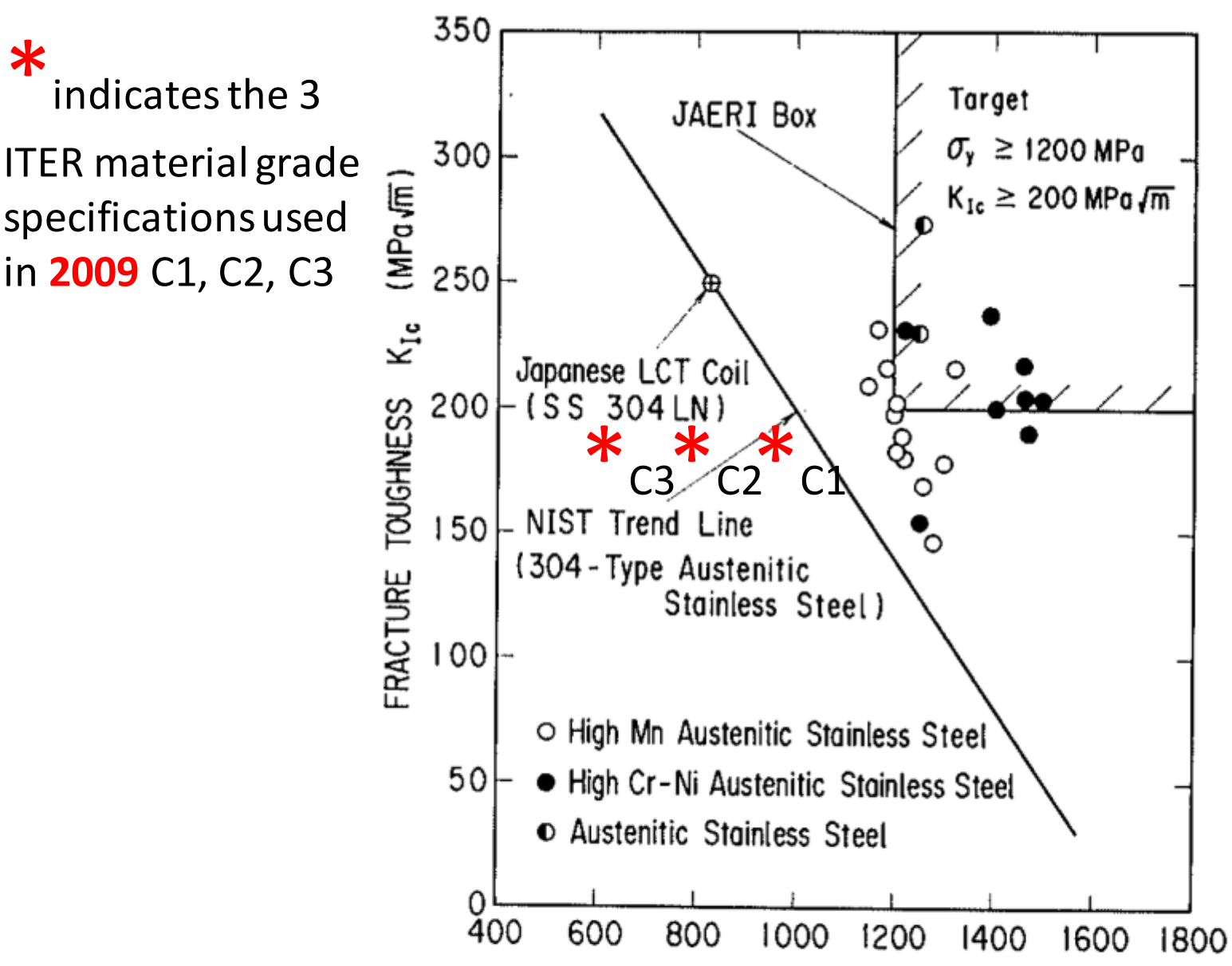

21. The relation between fracture toughness and yield strength of the JCS at 4 K. 1988

Table 1. Chemical compositions of the JCS.

| JCS | C | Si | Mn | P | S | Ni | Cr | Mo | N | Others |
|---|---|---|---|---|---|---|---|---|---|---|
| CSUS-JN1 | 0.026 | 0.99 | 4.2 | 0.026 | 0.002 | 14.74 | 24.2 | — | 0.34 | |
| CSUS-JKA1 | 0.023 | 0.42 | 0.49 | 0.006 | 0.001 | 14.0 | 25.0 | 0.68 | 0.268 | |
| CSUS-JN2 | 0.050 | 0.34 | 22.4 | 0.010 | 0.002 | 3.22 | 13.4 | 0.70 | 0.24 | V: 0.30 |
| CSUS-JK2 | 0.05 | 0.36 | 21.79 | 0.013 | 0.005 | 4.94 | 12.82 | — | 0.212 | Cu: 0.70 |
| CSUS-JJ1 | 0.046 | 0.44 | 9.74 | 0.020 | 0.002 | 11.92 | 12.21 | 4.89 | 0.203 | |

**Fig. 11:** Attempt to develop high strength steels in 1990s.

## 3.2 Conductor Jacket Structural Materials

The conductor jacket is also a way to include structural material in a magnet. However, to achieve the high field levels appropriate for a FPP (i.e. 12-14T), the majority of the coils (and those with the largest forces) require $Nb_3Sn$ and this in turn requires that the jacket material has to resist the $Nb_3Sn$ heat treatment (650C for about 200hrs) without degradation. The default material has always been 316LN but this suffers from degradation during the heat treatment. Starting in the 1990s, 4 alternatives have been considered:

1. Incoloy 908 susceptible to SAGBO (originating in US) [14]
2. JK2LB (a high manganese steel originating in Japan) [15]
3. Nitronic N50H (recently developed in China) [16]
4. Ultra low carbon 316LN [17].

Of these, Incoloy was used for the ITER model coils in the 1990s (and then dropped due to fabrication problems), JK2LB (a development of one of the high strength steels in Fig. 2, L meaning low carbon and B extra Boron) was used for the ITER CS coils. Ultra low carbon was used for all the TF coil thin jackets.

The development of Incoloy 908 was led by MIT, starting from considerations on requirements (in 1991). There was a perception that metal contraction coefficient from 600C to 4K should match that of $Nb_3Sn$ to avoid critical current degradation. This was subsequently shown to have been highly overestimated as an issue. Then in developing the alloy (a precipitation hardened nickel material), environmental issues were ignored. The material provided to be highly sensitive to Stress Accelerated Grain Boundary Oxidation (SAGBO) under which even tiny traces of oxygen in the heat treatment atmosphere led to catastrophic cracking under just the fabrication residual stresses, Fig. 12.

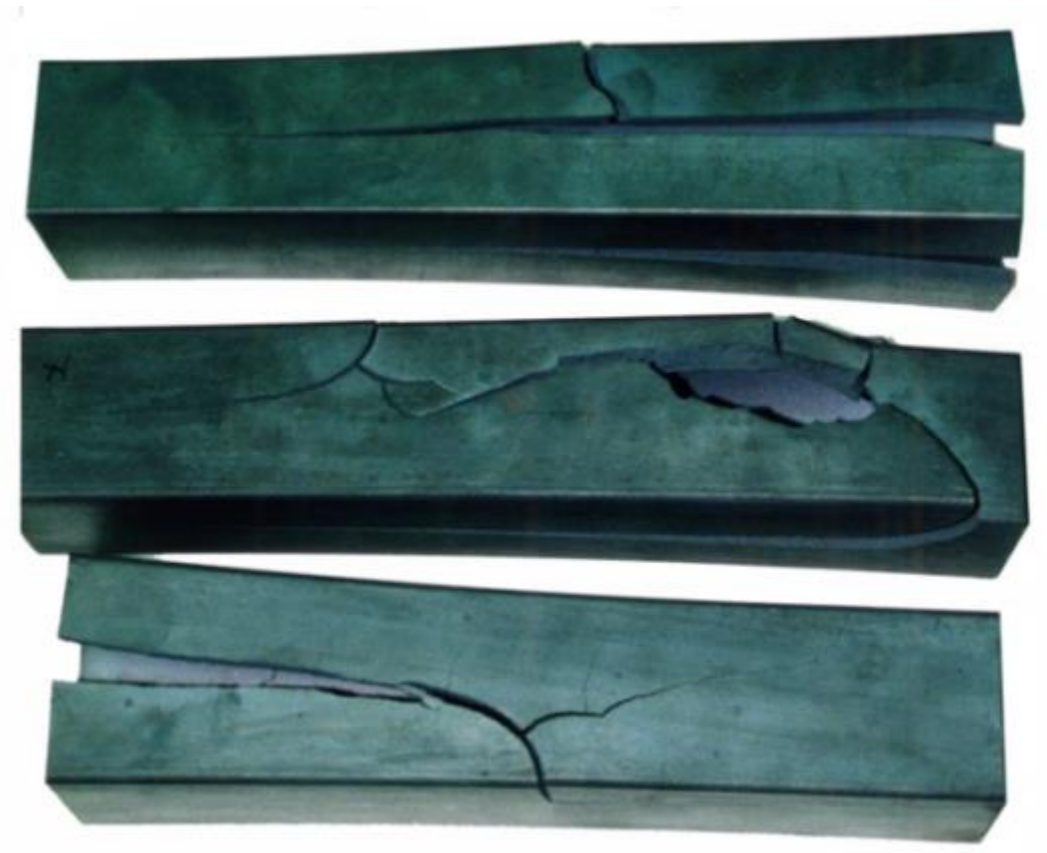

**Fig. 12:** Typical SAGBO cracking in Incoloy 908, in CS Model Coil jacket sections (K. Hamada and JAERI)

The development of JK2LB in the 2000s-2010s was more successful but again, in attempting to minimise the differential thermal contraction with the $Nb_3Sn$ strands, corrosion issues were overlooked. This time, the material was susceptible to halogen corrosion – halogens being normally used in soldering associated with the cable. The result was again cracking under residual stresses, Fig. 13.

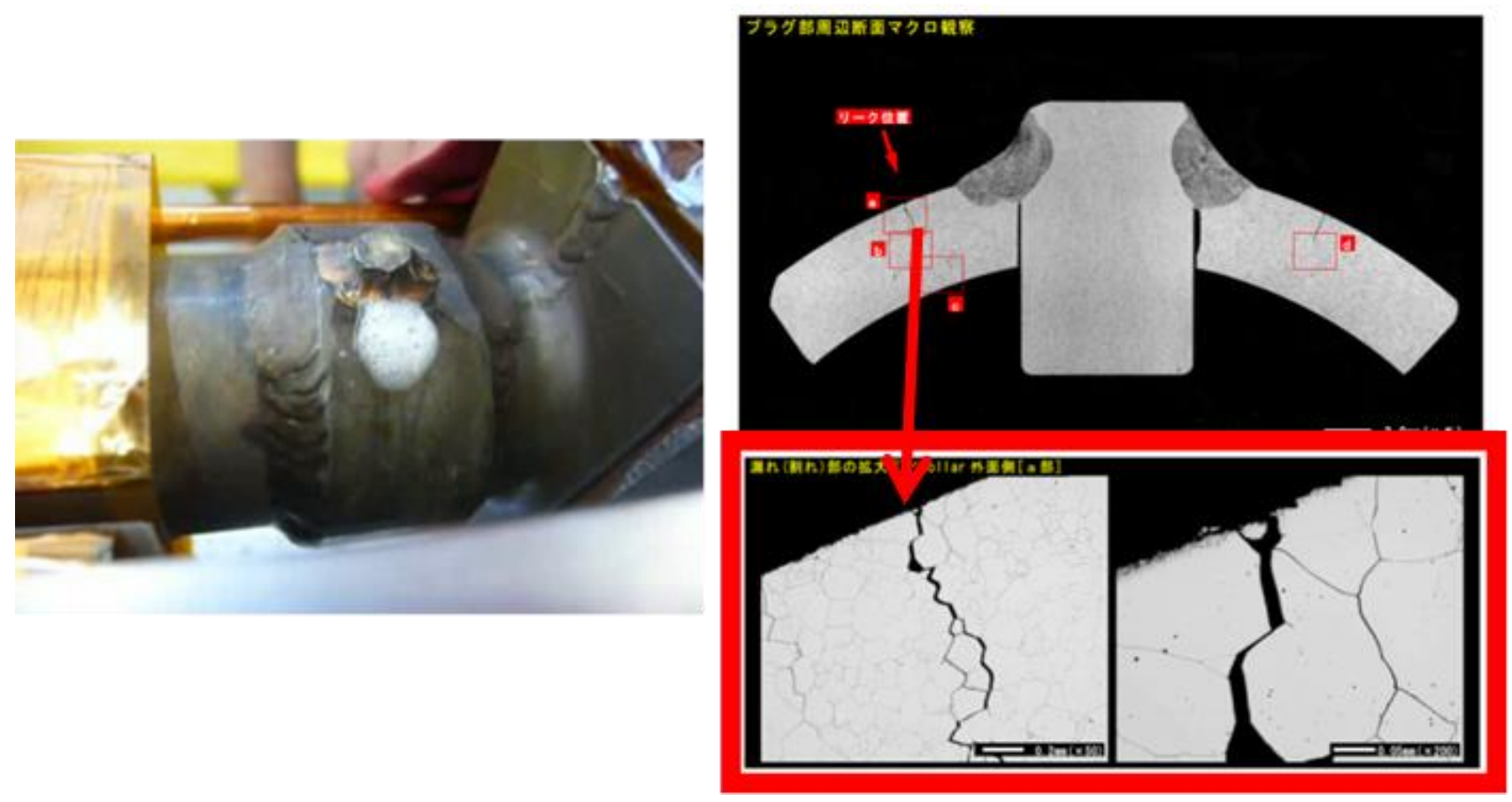

**Fig. 13:** CS JK2LB conductor samples 2012-13 - corrosion leaks originating from halides present in solder flux accidently contaminating the metal surface.

In both cases, coils were successfully built with the material but the risk of failure in a FPP was thought to be too high. A variant of Nitronic 50 has now been developed by ASIPP in China and may prove more successful. Also, the use of REBCO superconductors (at 12-14T) would relieve the problem of the $Nb_3Sn$ heat treatment (although the ASIPP variant of Nitronic 50 is claimed to resist degradation if exposed to it). [16]

## 3.3 Supply Chains and Supply Chain Technology

Stainless steels are a well known and broadly fully industrialised with long established producers. In the last 40 years the production capabilities for very high-quality steels for specific applications have increased, due to improvements in refining capabilities and downstream product formation, through to welding technology. These stretch from initial composition control, especially of trace elements during the ore/scrap reduction, inclusion elimination through refinement remelting, to specialised forging of high strength alloys without cracking. The upward trend, over the period of the ITER project, is shown in Fig. 14, and amounts to over a factor of 2. In overall production terms the FPP steel quantity in the magnets (about 30ktonne to match the scale units in Fig. 14) is negligible.

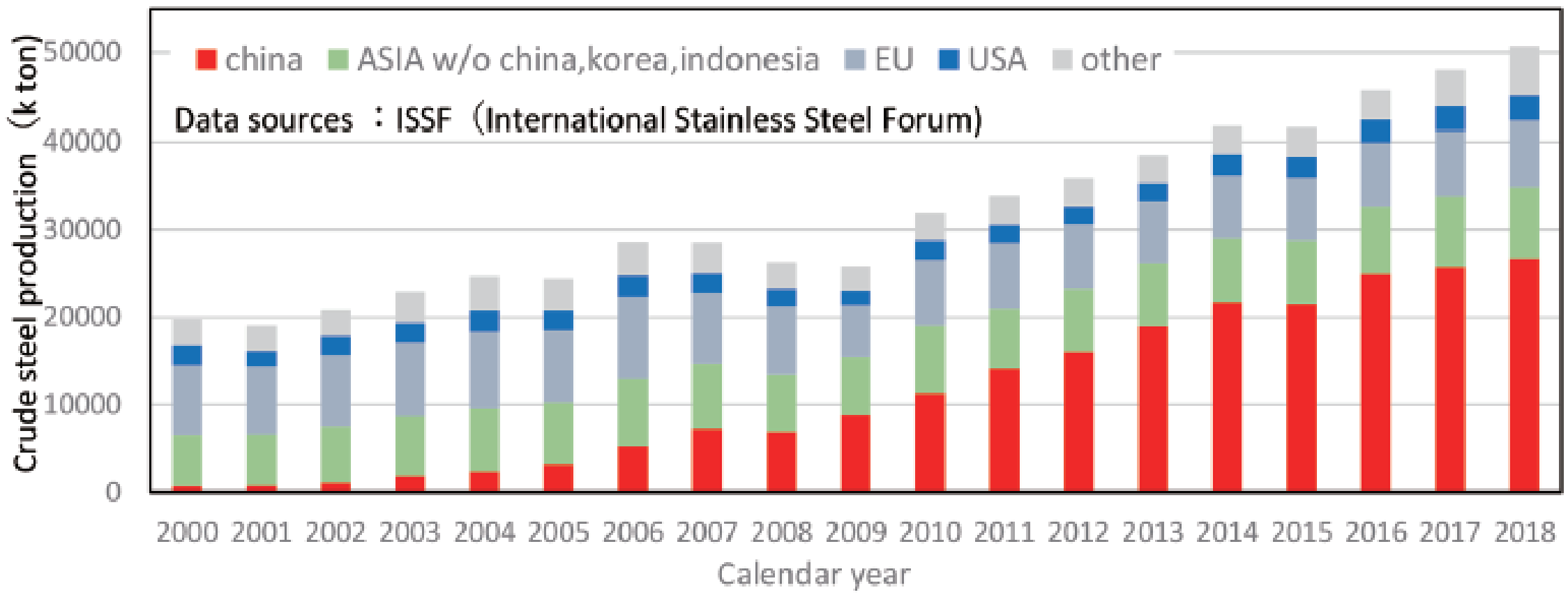

**Fig. 14:** World crude steel production of stainless steel, from [18].

### *3.3.1 Refining*

The background to the huge growth in the use of stainless steel is the improvements in structural performance and corrosion resistance that it offers. In this direction the magnetic fusion industry is so far a very minor customer. The problem is that although fusion is not competing against stainless steel for washing machines, it is competing, in similar materials and in particular in the similar product forms, against an expanding fission industry.

These improvements come substantially for the new developments, again over the life of the ITER project, in secondary steel refining, especially VOD, AOD and ESR (see definitions below). Initially the developments were in Europe, Japan and the US but since 2010 Chinese steel suppliers are leading.

From the late 1950s to 1960, many types of vacuum refining equipment were developed. Vacuum arc degassing (VAD) and vacuum oxygen decarburization (VOD) were developed in 1968, while argon oxygen decarburization (AOD) and ladle furnace degassing (LF) were developed in 1967 and 1971, respectively. These allow the reduction in Carbon content and reduce the risk of carbide precipitation and loss of fracture toughness especially for cryogenic steel applications.

One feature of the AOD and similar refining vessels that was not fully appreciated in the first decade of application was the ease of introduction and precise control of nitrogen as an alloy element in stainless steel. Initially considered only as an inexpensive austenitiser, nitrogen is now regularly used for its contributions to strength, corrosion resistance, and phase stability in austenitic stainless steel with high chromium and molybdenum.

Electroslag Remelting, Fig. 15, was initially (from 1964) carried out in the open air independently of the rest of the plant i.e. static crucible–single electrode or collar mould–electrode change. Difficulties in controlling oxygen and hydrogen in larger size ingots finally led to protective gas remelting from 1996 on. Specific plant concepts were developed for the production of high nitrogen steels—Pressure-ESR (PESR) and for the continuous production of small size billets at economic melt rates—Electroslag Rapid Remelting (ESRR). Due to the need for larger and larger forgings, ESR ingot sizes and weights have been increased up to 2600 mm ingot diameter and 260 t in weight (2016) [19].

What was an issue in 2001 for ITER (the low weight of ESR ingots and the lack of availability of ESR furnaces) had disappeared by 2015.

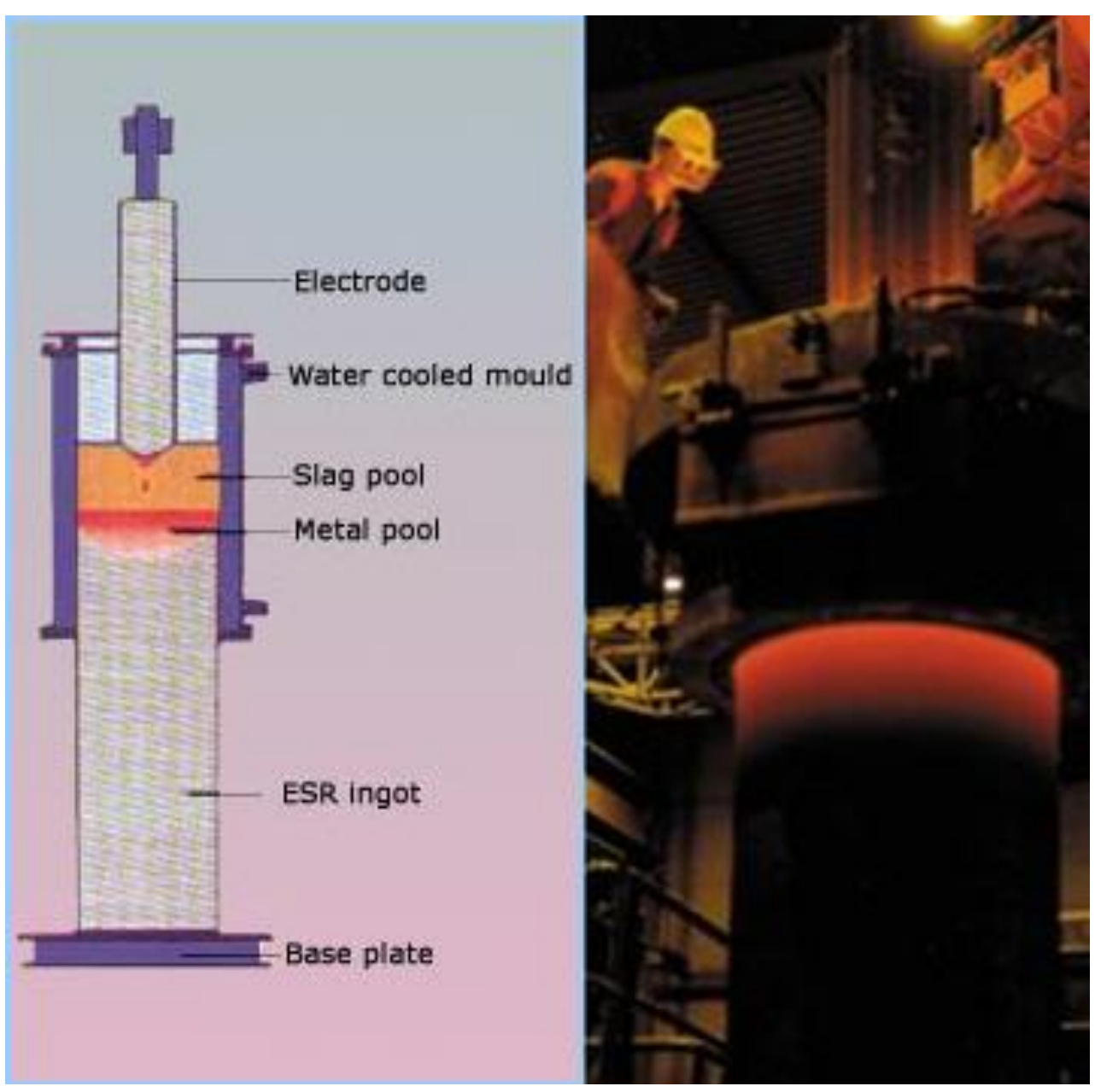

**Fig. 15:** Schematic representation of the ESR unit and actual furnace in operation [20].

### *3.3.2 Forming*

One of the ITER large projects initiated in the EDA was the production (within the TF model coil task) of large structural sections of TF cases. This work was partly completed after the end of the EDA in 2001 and is not well reported [21].

Figure 16 shows various early trials on forged sub-sections of the ITER TF coil case, showing the complexity of the forged forms. Top: seamless TF case (of course, cut to insert the winding but then with matching surfaces), bottom, seamless radial plate for TFMC and casting for TF outer intercoil structure.

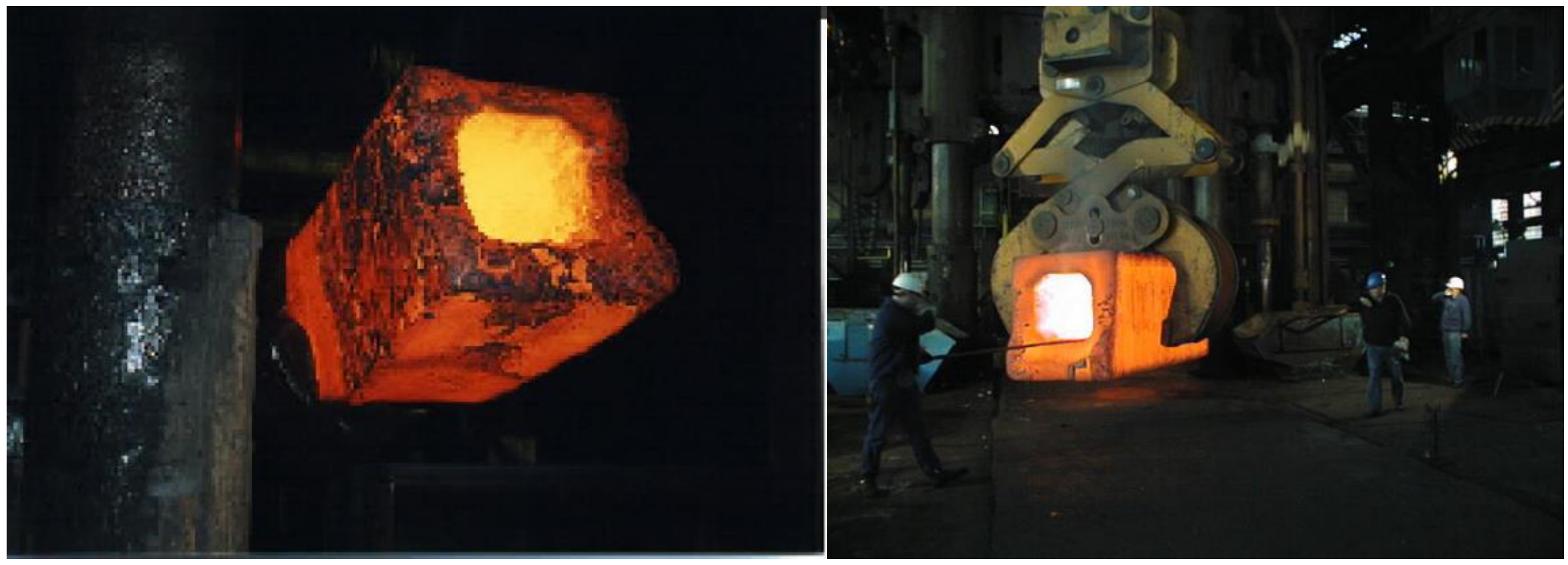

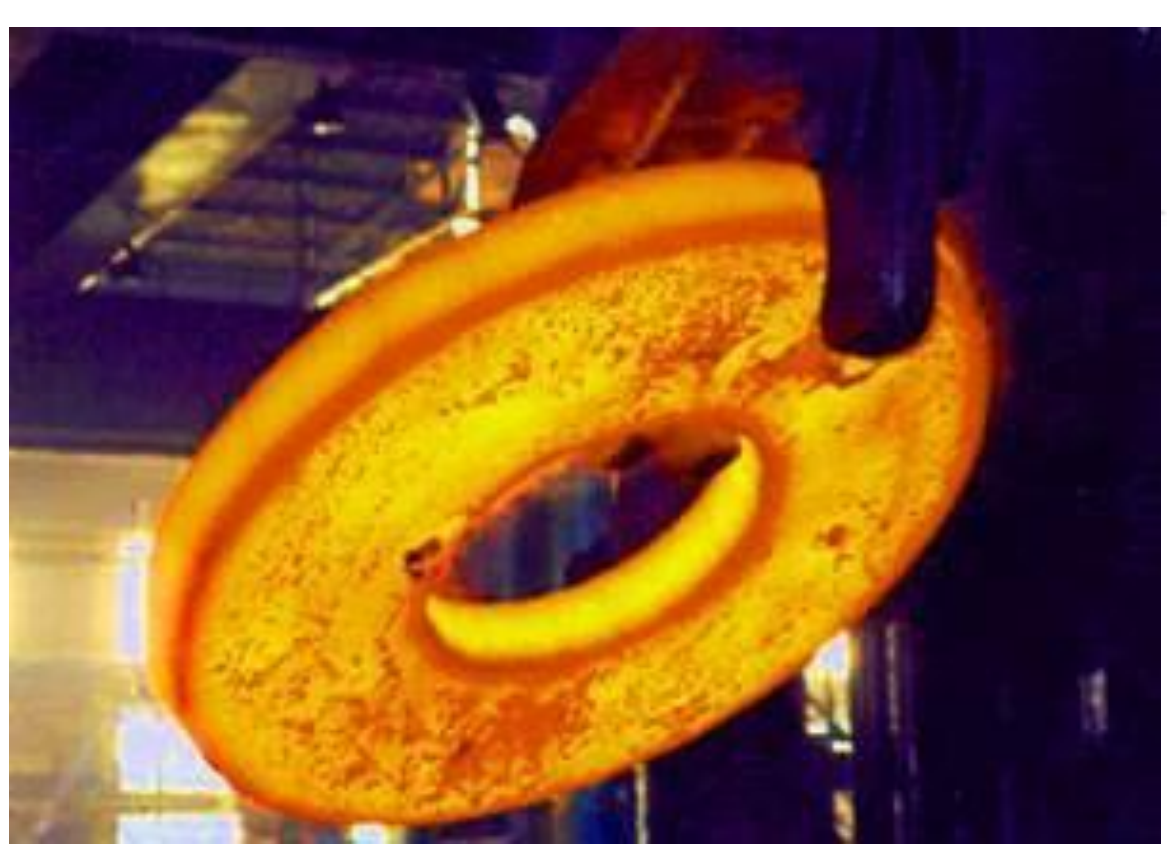

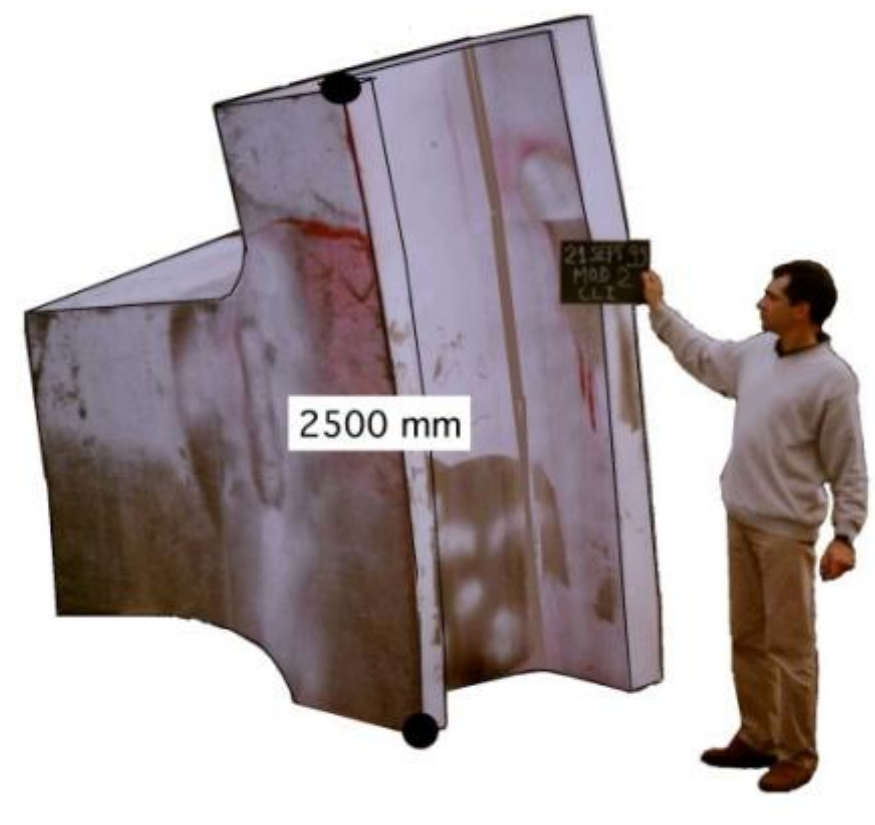

**Fig. 16:** Forging and Casting trials on Austenitic Steels, 2000-2005, in European Fusion Programme.

Casting was rejected because of poor properties (low modulus, low strength) and defects (voids) which were impossible to repair. The know-how obtained by the forging company (Kind) was used in the period 2012-2018 to produce a majority of the forgings for the TF coil cases and VV under contracts with EU, KO and JA.

As with other nuclear power plants, ITER relies on large amounts of forgings to achieve the best properties from a particular steel. One consideration for future optimisation could be the replacement of forgings by rolled plate, which is less specialised (and has a somewhat lower performance) but more readily available. The main problem is that plates are constant thickness, and flat or curved at constant radius. Building an ITER type case then requires inherently more welding with plates than with forgings, unless the design can be substantially simplified. The advantages of forging are being found in many areas, leading to rapid growth in the forging market, Fig. 17, and of course increased competition to access suppliers.

Responding to this demand, industrial forging capabilities have increased over the period of ITER manufacture. High grade austenitic stainless-steel supply in forged sections was a problem for ITER in the period 2010-2014, due to a limited number of suppliers capable and willing to supply it. Since then, forging technology has advanced with improved forces, more exact thermal control, robotic handling, induction heating and numerical process simulation. The difficulty for the small fusion sector is not the supply chain availability, but accessing expert suppliers with small orders and distant future growth prospects.

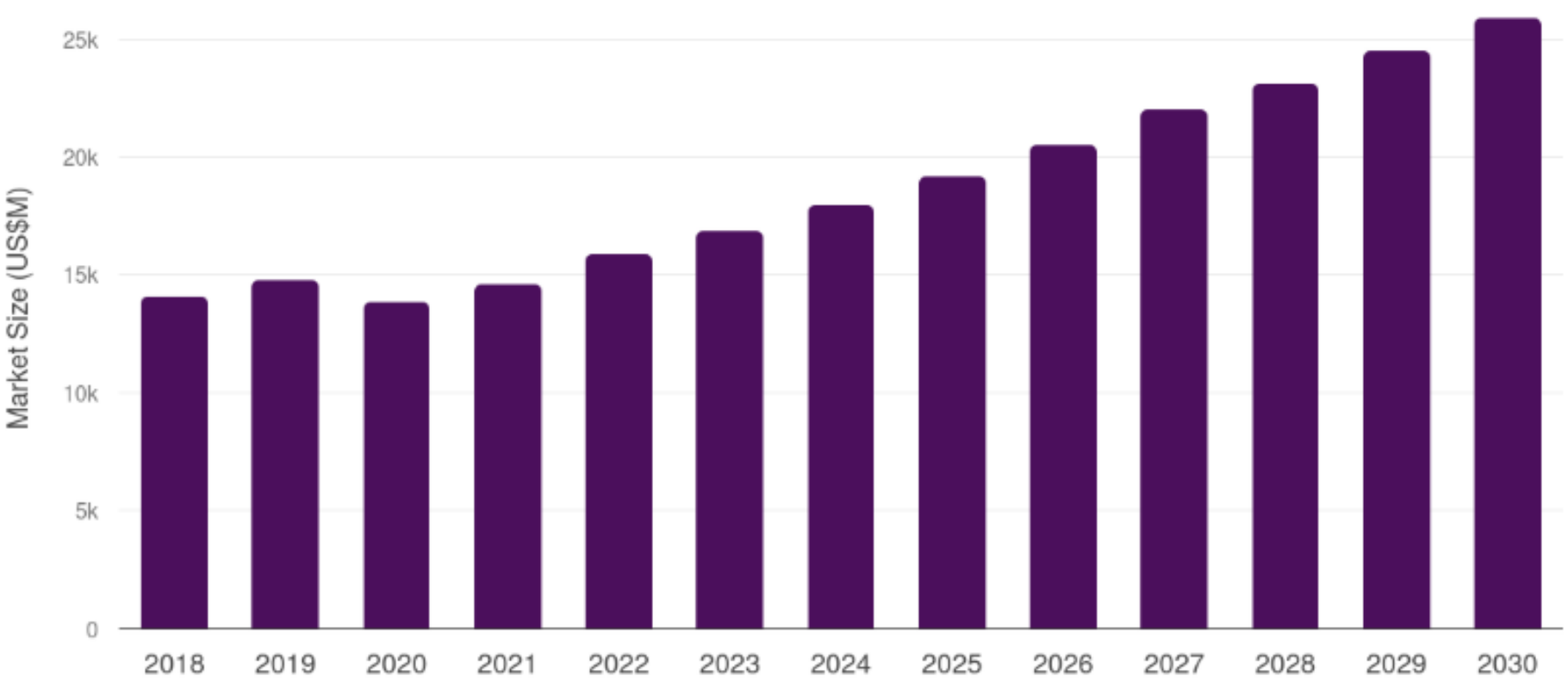

**Fig. 17:** The fastest growing segment is stainless steel [22].

### *3.3.3 Machining*

A FPP magnet structures are likely to require a very large amount of machining. If this is performed on subsections before welding (left picture in Fig. 18) the machine tools can be small. If it has to be performed on large final components (right picture in Fig. 18) then the availability of large gantry machine tools can be a bottleneck. In the case of ITER, in Europe the EU procurement agency supported the installation of such gantry machines at two suppliers for the TF coil radial plates [23]. The order, manufacturing and installation of such machines required >2 years and can be a major schedule issue, although it does ensure full access to the tooling throughout manufacturing.

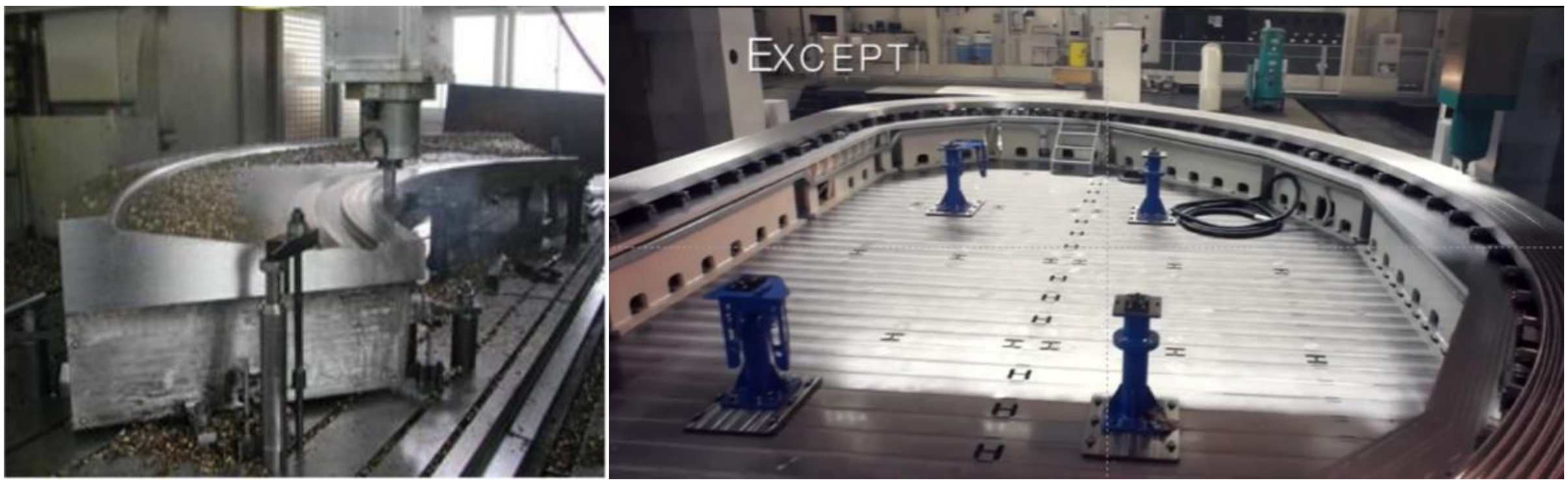

**Fig. 18:** ITER Case (Toshiba) and Radial Plate (SIMIC) Machining [23].

### *3.3.4 Welding*

Welding is invariably an issue with fabricating large dimension steel components. The dominant issue is weld distortion which, if it cannot be limited, requires extra machining steps and the addition of extra material through the manufacturing process. Distortion can be minimised using low heat input welding techniques like Electron Beam (also at low pressure with local vacuum) or Laser. The main weld quality issues are given in Fig. 19. These are not generally insurmountable obstacles but issues to be solved during the structural manufacturing design and qualification with an appropriate program.

| Welding Issue | Primary Structural Hazard | Industry Standard Prevention |
|---|---|---|
| **Severe Distortion** | Geometric misalignment; fit-up issues. | Use rigid fixtures, balanced symmetrical welding sequences, and low heat welding (Electron Beam and Laser) |
| **Hot Cracking** | Immediate structural tearing in multi-pass welds. | Maintain a 4%–12% Ferrite Number (FN) |
| **Sensitization** | Rapid localized corrosion (weld decay). | Use ultra-low carbon base metals (L-grades), lower the heat input |
| **Nitrogen** | Voids formed in the weld. | Use manganese based fillers |

**Fig. 19:** Typical Austenitic Steel Welding Issues.

Figure 20 shows the semi automatic welding assembly of an ITER TF coil case [24]. The availability of skilled welders was also an issue for the ITER project. Fully automatic welding is preferable but generally requires a better alignment of the weld edges which can be very difficult to achieve on large components.

Welding of ITER structures used narrow gap TIG, Laser and low pressure EB (sometimes in combination). Due to different supplier preferences, different methods were sometimes applied to the same welding step in different coils.

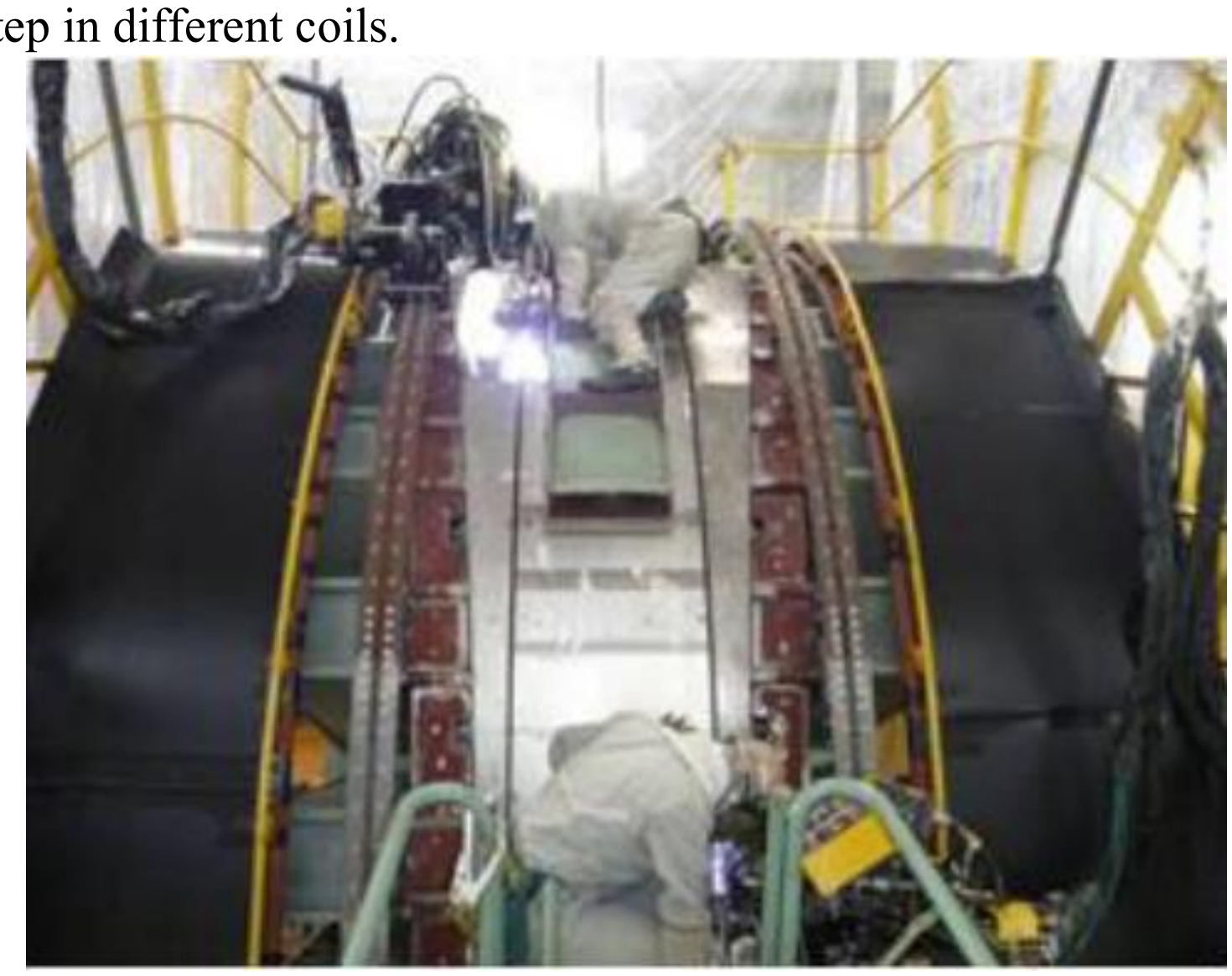

**Fig. 20:** Welding (semi automated) of ITER Case, Toshiba, Fabrication process of A1 segment full-scale mock-up [24].

## 4 Structural Performance Assessment for Magnet Structures

### 4.1 Codes, Standards and Criteria

Codes, Standards and Criteria (CSC) are important as they are a way of implementing lessons learned from previous experience. They define how we have to make the product (in this case a steel structure) so that it has a similar performance to the one we made before. The question for magnet structures for FPP is which ones to apply, or whether to develop our own.

A well known CSC example is the ASME Pressure Vessel Code, from the 1900s. It originated from a series of disastrous boiler explosions and it defined (1) what materials could be used in boilers (2) how they can be formed and joined (3) wall and flange and nozzle thicknesses matching the pressure based on standardised analysis route and safety margins (4) ways to check the overall processes (up to and including pressure proof tests). It became a legal requirement, all based on experience and diagnosis of the reasons for failures (as known 120 years ago) [25].

A side note at this point is that knowledge of performance is gained from failures and the unexpected. Often we see First-of-a-Kind (FoK) magnets tested and the results recorded as 'performed according to expectations', and a box is ticked. Possibly the manufactures were extremely able and foresighted. Or maybe they didn't look hard enough….and will be caught when some shade of the operating conditions changes and provokes an unforeseen condition,

A further note is that it is difficult to write Codes or Criteria without experience, although this is being proposed for FPPs. Codes and standards are based on lessons learned. With FoK FPP there are none. It is possible to look for similarities from other components and chose Codes and Criteria (C&C) on this basis…good for a start but by far not the end of the story. But it is not possible to assume for FPP that just because a set of codes and criteria are followed, there will be adequate margins. C&C applicability has to be justified as part of the licensing process and this must come from their existing application to a FPP, not from some different component. In particular, superconducting magnets differ substantially from pressure vessels and despite developing sophistication in understanding failure mechanisms in structures, 'bolting' magnet design codes onto pressure vessel design codes appears risky and will hinder development of original magnet design codes that will allow better optimisation. FPP Codes and Criteria are under active development, with some attempting 'bolting' actions for magnets [26].

ASME rules for the construction of fusion power plants are being established under Section III, Division 4 of the ASME Boiler and Pressure Vessel Code [27].

Almost every engineering project in 2025 will have in addition to C&C, hundreds of applicable standards and standardised procedures, and there are multiple families (ASME, ASTM, ISO, EN for mechanical) and fields (mechanical, electrical, hydraulic, construction, machinery etc etc) and environments (space, aeronautic, cryogenic, underwater etc etc) and materials (metals, composites, ceramics etc etc). Many of these standards and procedure refer to processes (a typical example is welding) and unlike C&C, are applicable to magnet structures.

For magnets, C&C are dominated by fabrication issues and imperfections (which are of course themselves fabrication issues), i.e. exactly the things needed to be controlled ensure continuity of performance. It was clear in the very early stages of ITER (i.e. late 1980s and early 1990s) that there are features of cryogenic large magnets that differ substantially from components sometimes seen as similar, large nuclear pressure vessels.

- The magnets operate 4K-77K, mostly 4K. Compared to high temperature, yield and ultimate strength are increased but fracture toughness similar. Fracture is relatively much more important as a failure mode than plastic yielding.
- Fracture depends on peak stress (concentrations, residual etc). Process of stress linearization no longer valid. The fabrication history (creating residual stresses) is important
- In-service inspection of the magnets is generally limited, operational monitoring of structural behaviour (displacements, for example) is insensitive (although electrical condition monitoring may be more sophisticated).
- Leak-before-break is not generally a design option and in some cases fast fracture may occur. In this respect magnet structures are different from nuclear power plants
- Extensive use of non-metallic materials especially for bonding and compressive load transmission.

- The loads can have strong cyclic components, not similar to Nuclear Fission (although some similarities to aeronautical)
- Electromagnetic loads cause 3D stress systems, generally need FE analysis, far different to pressure vessels.
- Combination of structural support functions in a High Voltage environment; the structural limits can be defined by the electrical functionality.
- Yield stress is increased above level of residual fabrication stresses. Limiting of stress peaks through local plasticity is less extensive than at room temperature and above, so stress relaxation through ratchetting (basis of high temperature yield stress margins and primary/secondary stress separation) does not occur to same extent
- Limited previous experience with such a design and manufacturing technologies/ materials are evolving.

Pressure vessel codes in the 1990s and early 2000s, and especially nuclear ones like ASME (and to a lesser extent RCC-MR [28]) provide a 'cradle to grave' prescription for Pressure Vessels and (by extension) to (mainly) PWR type fission reactors

ASME originally (i.e. in 1970s) prescribed materials, product forms, allowable joining techniques (i.e. welding), analysis methods (based on formulae), safety margins and in-service inspection. Implicit within the codes are assumptions about material behaviour (ductile, above room temperature, basically static). They date from a time when metallic materials could only be assessed in the linear elastic regime, and as defect free. Imperfections were screened out by fixed criteria and a high level of conservatism.

The ASME BPVC for nuclear vessels dates from 1963, when ASME published the first edition of ASME BPVC, Section III, "Nuclear Vessels". Over the years, the scope of ASME BPVC Section III has expanded to cover practically all of the pressure and liquid storage components involved at a nuclear power site. A further section of the ASME BPVC that we refer to is Section XI, which began in the 1980s as a set of rules for in service inspection of the primary pressure boundary system of nuclear power plants. It has evolved to include other aspects of maintaining the structural integrity of safety class pressure boundaries. These include procedures for component repair/replacement activities, analysis of revised and new plant operating conditions, and specialized provisions for non-destructive examination of components and piping. Since the 1990s a further in-service inspection code has been developed, API 579 (becoming ASME FFS-1 (Fitness for Service) in 2007). [29]

- ASME Section XI: Provides mandatory regulatory rules for periodic non-destructive examination (NDE) schedules, baseline tracking, system leak tests, and specific nuclear repair administrative controls.
- ASME FFS-1: Provides quantitative engineering evaluation techniques (using a tiered 3-level approach from basic screening to advanced finite element analysis) to decide if degraded or damaged equipment (corroded, cracked, or fire-damaged) can continue safe operation ASME FFS-1 has developed now (2021) into a code for assessing structures with imperfections.

The specific features of ITER Magnets drove the need for specific set of mechanical design criteria for magnets. The first efforts appeared in the ITER EDA (in 1990) and ITER decided (already before 2001) to use a 'defect based' code for the metallic mechanical design criteria on and essentially certify the magnets for a defined fatigue life based on the results of the manufacturing inspections and the structural analysis. Once the life is exhausted, the magnets should shut down unless some form of in service inspection can be performed. In effect the ASME Section III prescription of perfect structures is dropped in favour of imperfection and an obligation to quantify how the imperfections develop into failures in order to provide a certified lifetime. This procedure can exploit several of the procedures defined in API-579 /ASME FFS-1, as summarised in [30]

This methodology defines an allowable operating life based on the inspection levels set for manufacturing. By the FDR in 2001 there was a full version of the structural design criteria available which focused on fatigue and fracture and expanding the concept of a defect-based design, based on the Fitness for Service assessment method set out in API 579, later (in 2007) adopted into ASME as ASME FFS-1. This had important consequences for manufacturing since allowable weld defects can be set according to the total stress levels, making for less stringent NDT and fewer repairs (see Section 4.4). The ITER design criteria for magnets are available only as internal references, [31].

ASME and other codes have been gradually updated in the 2000s to make them more widely applicable, and attempts have been made (especially in JSME) to introduce components for magnets. These need to be treated very cautiously as they still extend fission philosophy into fusion and lack fusion magnet design input.

Gradually the 'traditional' prescriptive part of ASME is being relaxed and the code is including a 'defect based' approach which includes methods to model failure (the simplest being elasto-plastic analysis to establish limit loads). Fatigue by LEFM assessment is now included. RCC-MR is following a similar trend.

However, some weaknesses still remain, see the analysis section.

### 4.2 Structural Performance Assessment

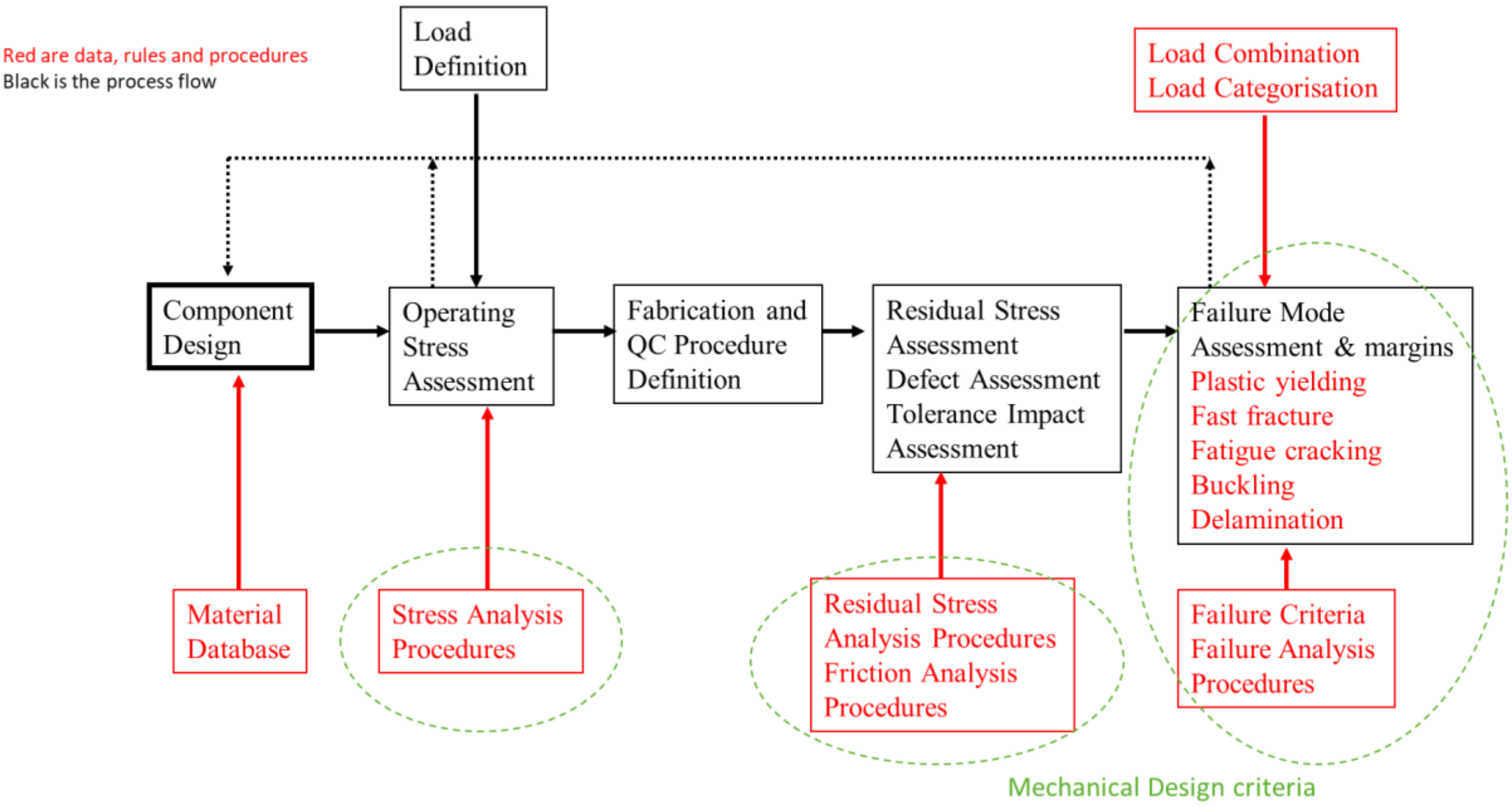

4

**Fig. 21:** Structural Assessment Diagram.

The Structural Assessment procedure for the magnets is outlined in Fig. 21. Considerable detailed knowledge is required of the manufacturing process and the resultant stresses. Originally (in 2001) approximations (semi-analytical expressions or fits to simplified finite element models) were used for these, as given in API-579 and procedures such as [32].

Over the long period of ITER magnet design and construction, there have been major changes in structural analysis procedures. At the time of the 2001 FDR most structural analysis was linear elastic, with inclusion of gaps and sliding (and limited model sizes). Now, multiphysics models are generally used, highly non linear and capable to simulated manufacturing processes (like welding and welding stresses) and failure events like fracture from cracks and plastic yielding.

Despite these advances the basic method of structural assessment, calculation methods of stresses to apply in the design criteria are still those of the 1980s and 90s. So sophisticated techniques are used to find the stresses and these are 'linearised', being converted into Membrane and Bending Stresses, Primary and Secondary, and Peak (i.e. the terminology of analytical based structural analysis of the 1970s). Adaption in the 80s and 90s acknowledged the role of numerical analysis but meshes are this time were coarse and tended to create numerical stress concentrations when used with linear elastic models.

The only apparent benefit from this process is to be able to use safety factors developed in the pre-numerical era at which time the analysis models conveniently gave primary, bending, secondary stresses separately and didn't show stress peaks.

The combined limits on primary and secondary stresses are a way of saying that 'settlement' (work hardening) will occur over a few load cycles and the peak stresses are reduced (and a residual stress system is created) and are therefore acceptable [33].

This approach to stress assessment shows limitations and inconsistencies when it now includes Linear Elastic Fracture Mechanics (LEFM) and fracture assessment, which include peak stresses and residual stresses (which can really only be evaluated by FE analysis and often with plasticity included). ASME now recommends elasto-plastic analysis for complex geometries without however saying what the acceptable stresses are….implicitly the analysis must cover a few load cycles to see if stress stability is reached (in which case, the allowable design stress intensity Sm ceases to have any meaning and residual stresses become built in).

It is becoming timely to update the magnet design codes used for ITER to match modern analysis techniques and reduce margins. We don't need better materials but do need to exploit better the ones we have, use better material quality controls and inspection methods, and free ourselves from the historical baggage of an association between magnets and pressure vessels. Although not included in this paper, there are similar opportunities to improve the failure modelling of bonded components and resins which are frequently an integral part of magnet windings and where electrical considerations are also needed. ITER developed design criteria for these as well, which could also be advanced to the next generation [34].

### 4.3 Main Loads

Magnetic loads in FPP are extremely large as well as being multi-directional. A simple circular magnet in isolation has some similarities to a pressure vessel. There are no net forces and the magnetic loads act as if to burst the circle, as with internal pressure.

This similarity disappears once we consider complete magnet systems for fusion devices. Taking ITER (a tokamak) as an example, Fig. 22 shows the resultant loads on individual coil systems and Fig. 23 the time dependent out of plane loads on individual TF coils (which have to be reacted through the intercoil structures). The structures have to react a complex 3D force pattern and the resultant force magnitudes are huge. In plane force on each TF coil is 40200t, and the upper and lower parts of CS apply 50000t towards each other at the centre.

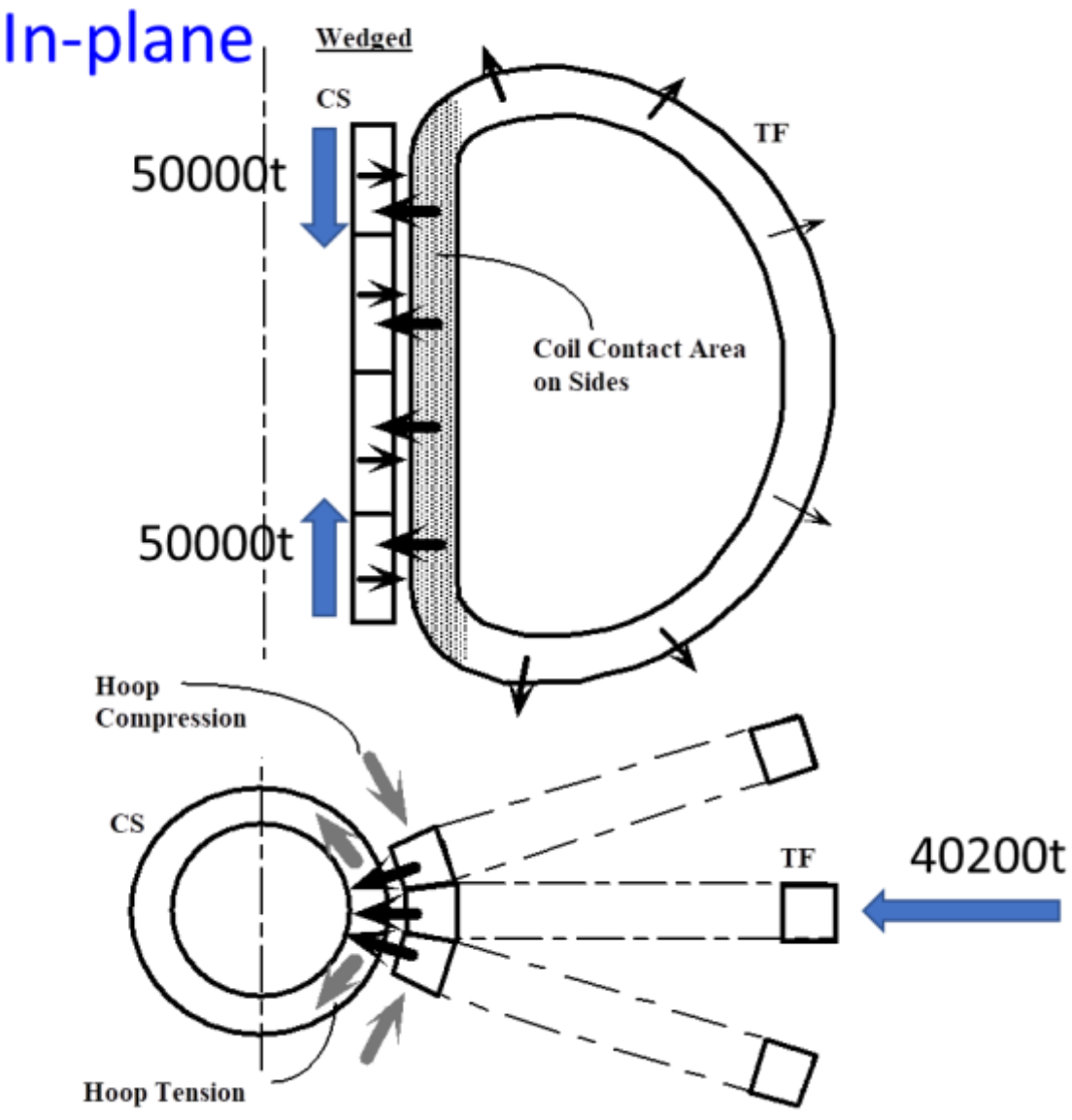

**Fig. 22:** Main Resultant Loads on ITER Magnets.

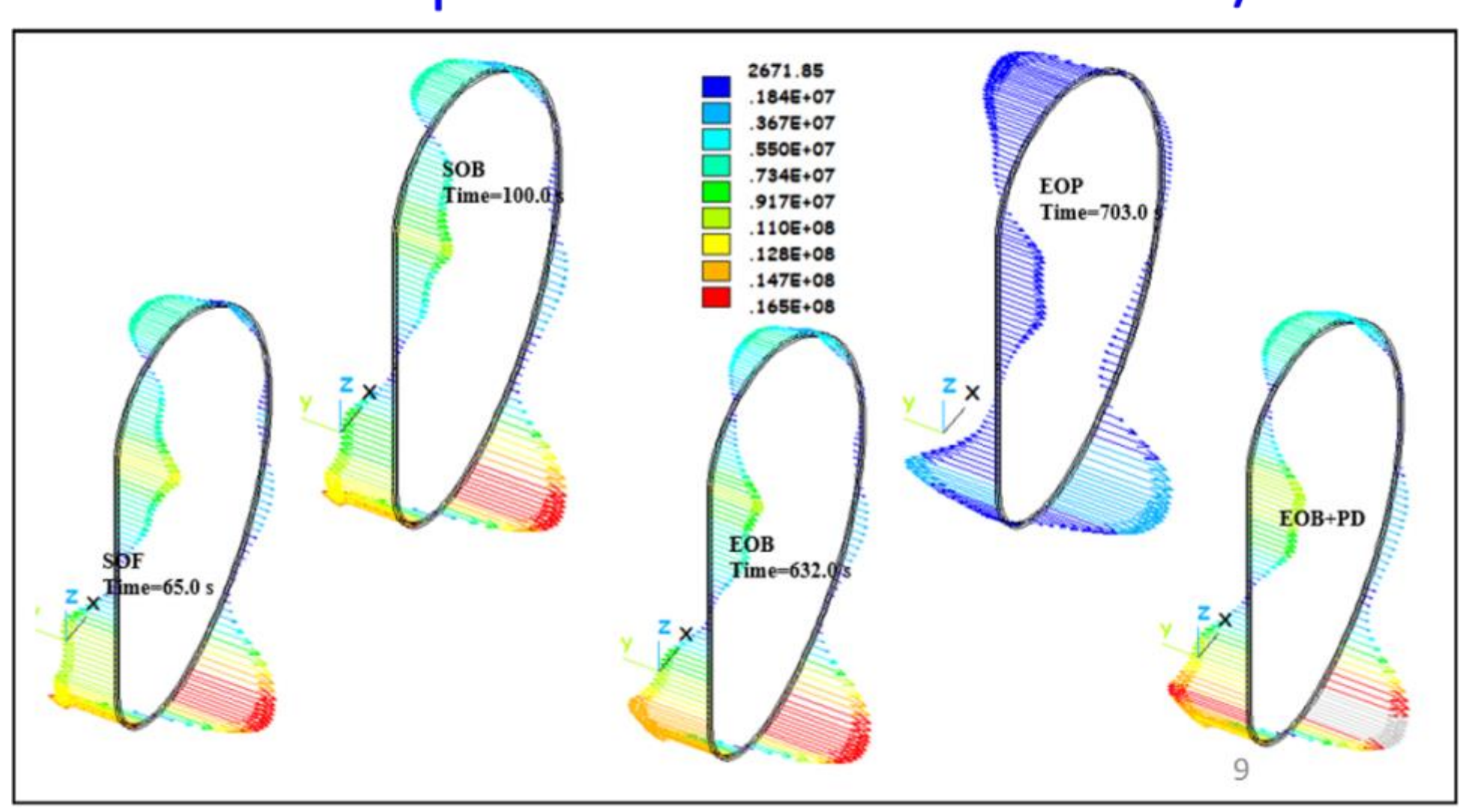

**Fig. 23:** Out of Plane Loads (due to interaction with the poloidal field) on a TF magnets.

### 4.4 Analysis Methods

There is a vast amount of structural analysis of the ITER magnets from the 1990s until the 2020s, evolving in sophistication and complexity according to finite element analysis capability, as well as computational power. Two examples have been selected:

#### *4.4.1 Case Stress Concentrations, Fig. 24*

This shows the 3 stress concentrations in the bulk case. All three are coincident with manufacturing welds, and proper structural assessment is based on LEFM and requires calculations (or estimations) of the residual weld stresses.

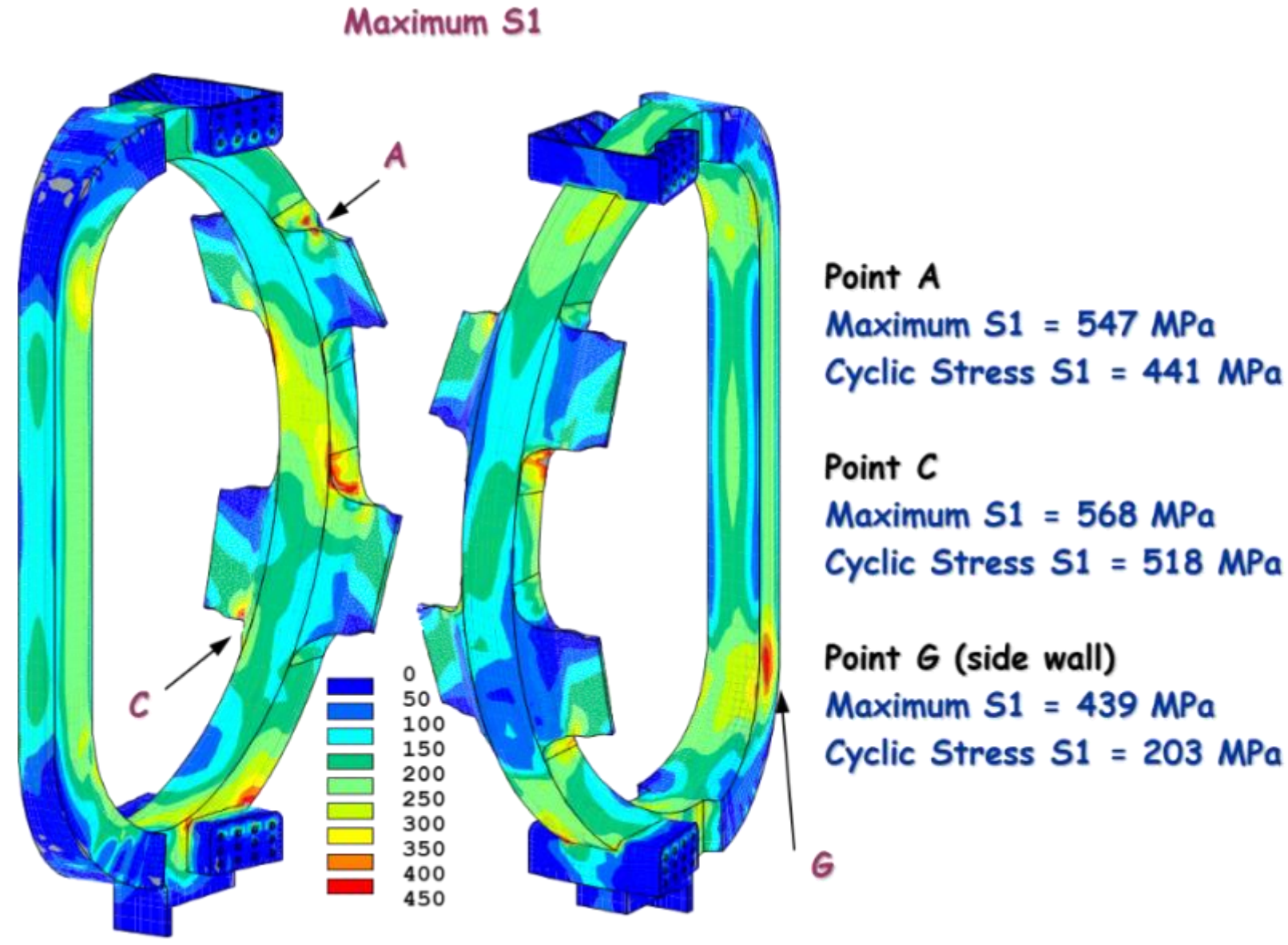

**Fig. 24:** Maximum Principal Tensile Stress in the TF Coil Case.

Figure 24 also illustrated the relative inefficiency of the case design, with large volumes at very low stresses (i.e. far from the stress limits) and a few peaked areas.

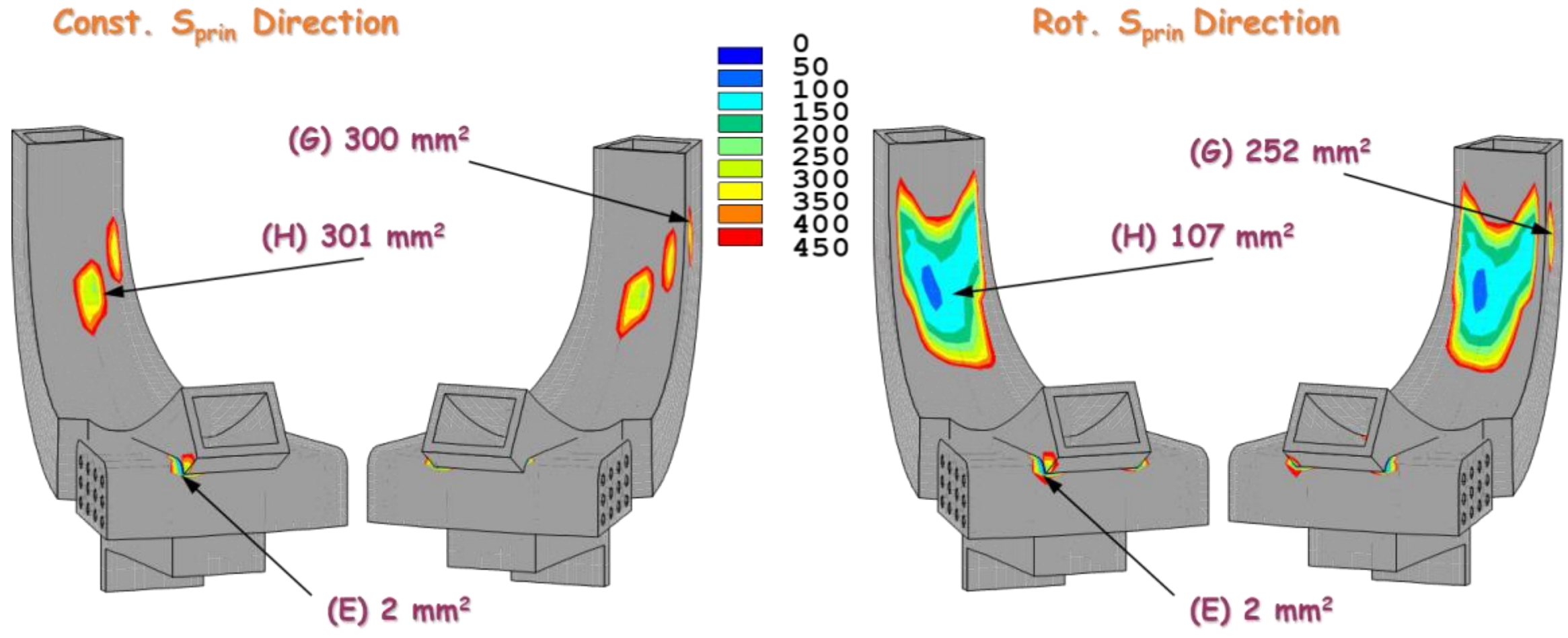

**Fig. 25:** Example I of LEFM Fatigue Assessment on Magnet Case to Determine Allowable Initial Defect.

#### *4.4.2 Case Defect Size Assessment Figs. 25, 26*

In Figures 25 and 26, a fatigue assessment according to the FFS LEFM procedure is carried out to provide a value for the maximum initial (t=0) allowable subsurface defect to sustain 60,000 cycles (safety factor of 2 on cycles). In the fatigue analysis assumptions are made about residual stresses in base metal and welds [35]. The NDT inspection has to be calibrated to detect defects at least as small as this

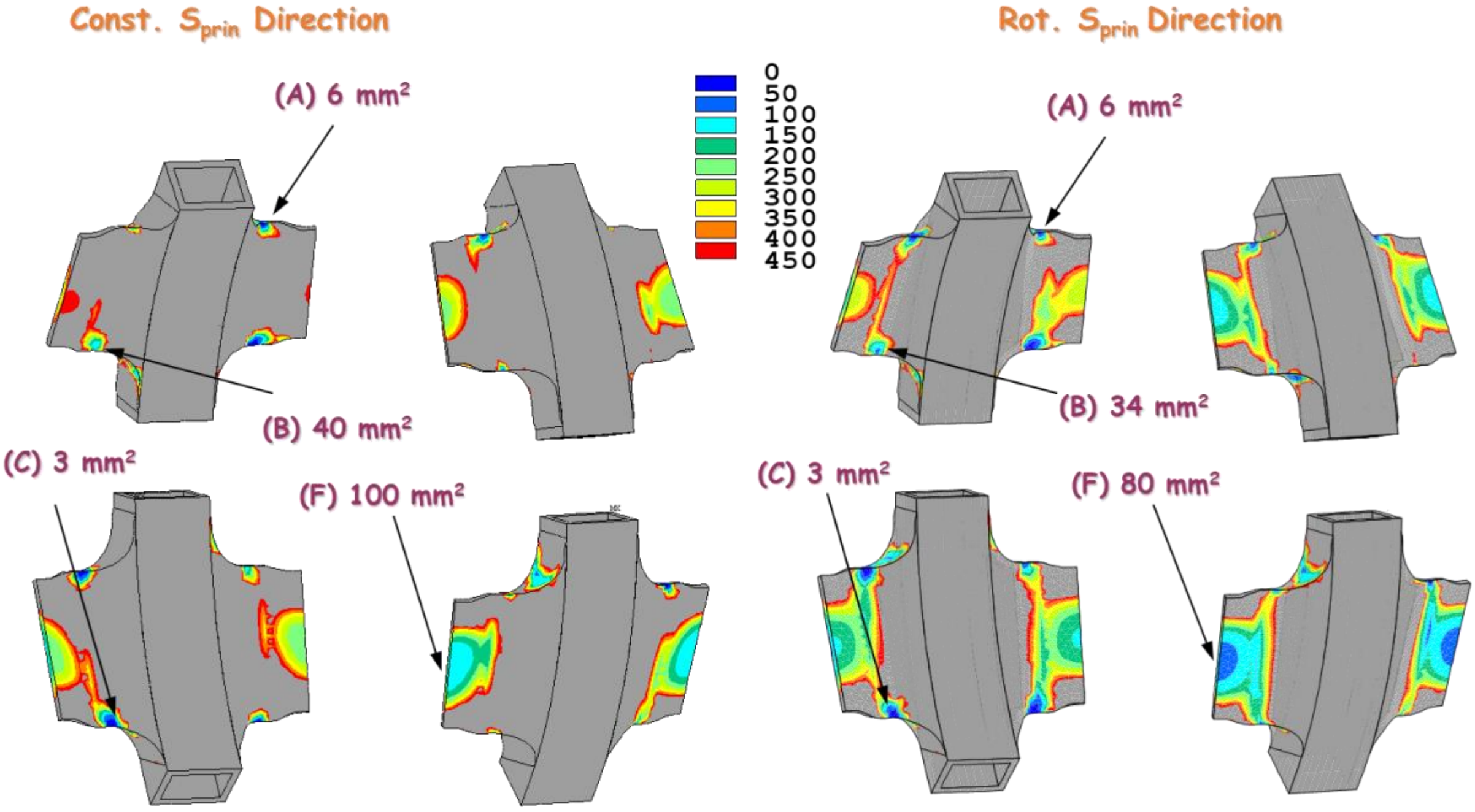

**Fig. 26:** Example II of LEFM Fatigue Assessment on Magnet Case to Determine Allowable Initial Defect.

These analyses are all non linear and include sliding and gaps at interfaces. Analyses from 2015 included welding simulations, elasto-plastic evaluation of stress concentrations and crack tip stress intensity evaluations in a 3D stress field.

## 5 Structural Design for Cryogenic Metallic Structures

Structural design is usually interpreted to mean obtaining acceptable stresses using finite element analysis. However, to match more sophisticated design criteria involving failure assessments, a much more sophisticated approach is required, integrating also the manufacturing route(s). This was difficult to achieve in the ITER environment since the design needs to match the manufacturing route and this was not known until the Dometic Agencies (the ITER stakeholders) tried to place contracts. The magnet structures use a defect-based design and defects are associated with welds. Minor areas can dominate performance, with large volumes not near critical stresses, and critical stresses in local areas well below those achieved in the base material. These areas are typically welds but not just weld defects but also the achievable weld layout. It was found in 2015 that the structure supplier could not (at reasonable cost) achieve full penetration welds in many of the structural attachments, and such welds must be treated as a defect according to the codes. Accordingly, an assessment procedure had to be developed for such welds [36].

### 5.1 Subpart Breakdown

The ITER magnets were not very well designed as regards coordinating manufacturing limitations with mechanical design criteria. They were essentially designed by structural analysts, with poor consideration of the manufacturing constraints (shaping, joining and inspection) and overall, a very inefficient utilisation of material. Figure 27 shows the huge number of attachments and assembly welds in a single coil case. Figure 28 shows the main segments under manufacture, with a lack of consideration of worker access and weld angle resulting in multiple time-consuming movements of the subsectors during welding and eventually a need to relax the requirement for full penetration welds on many of the flanges.

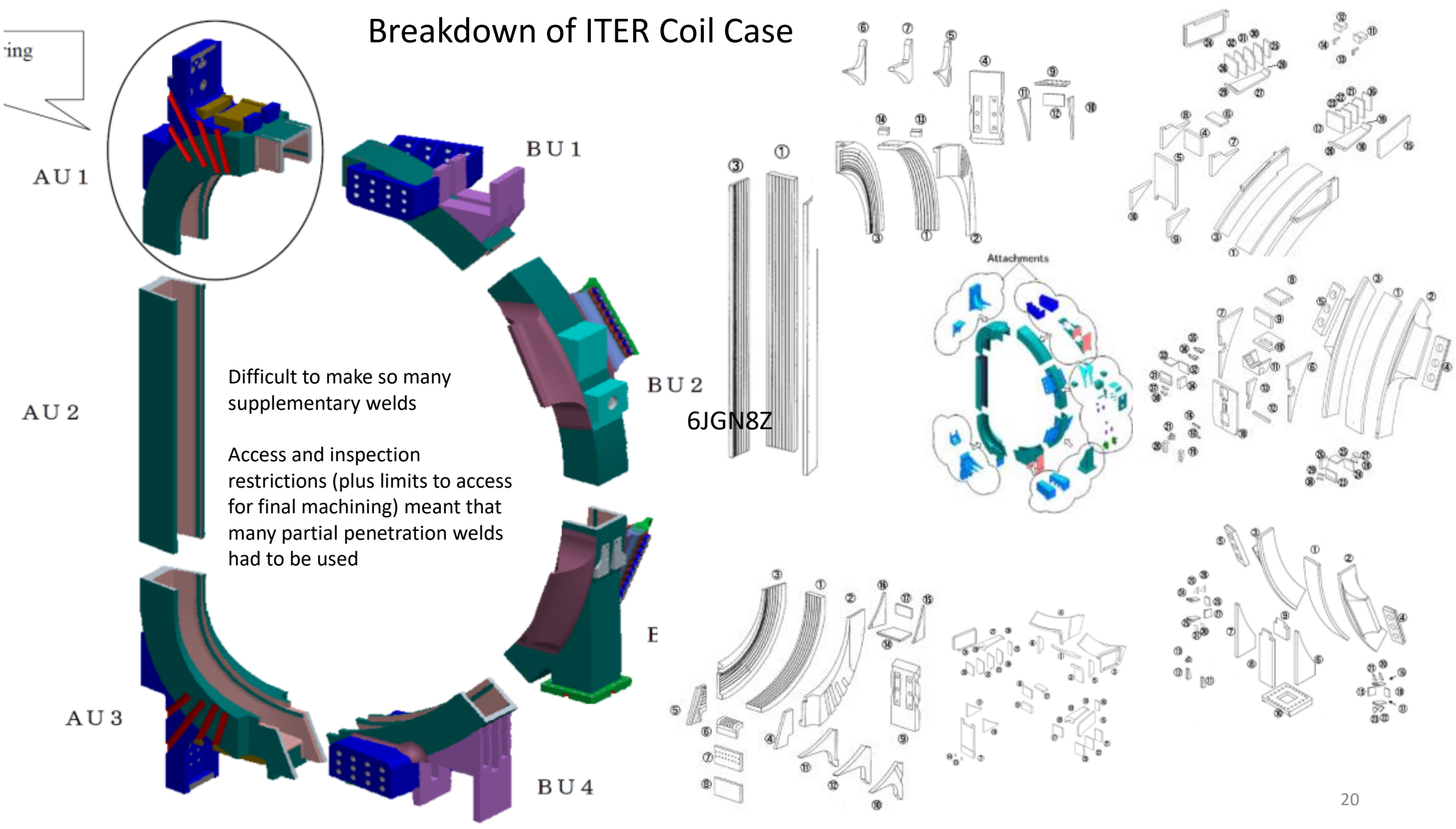

**Fig. 27:** Sub-components making up a TF coil case.

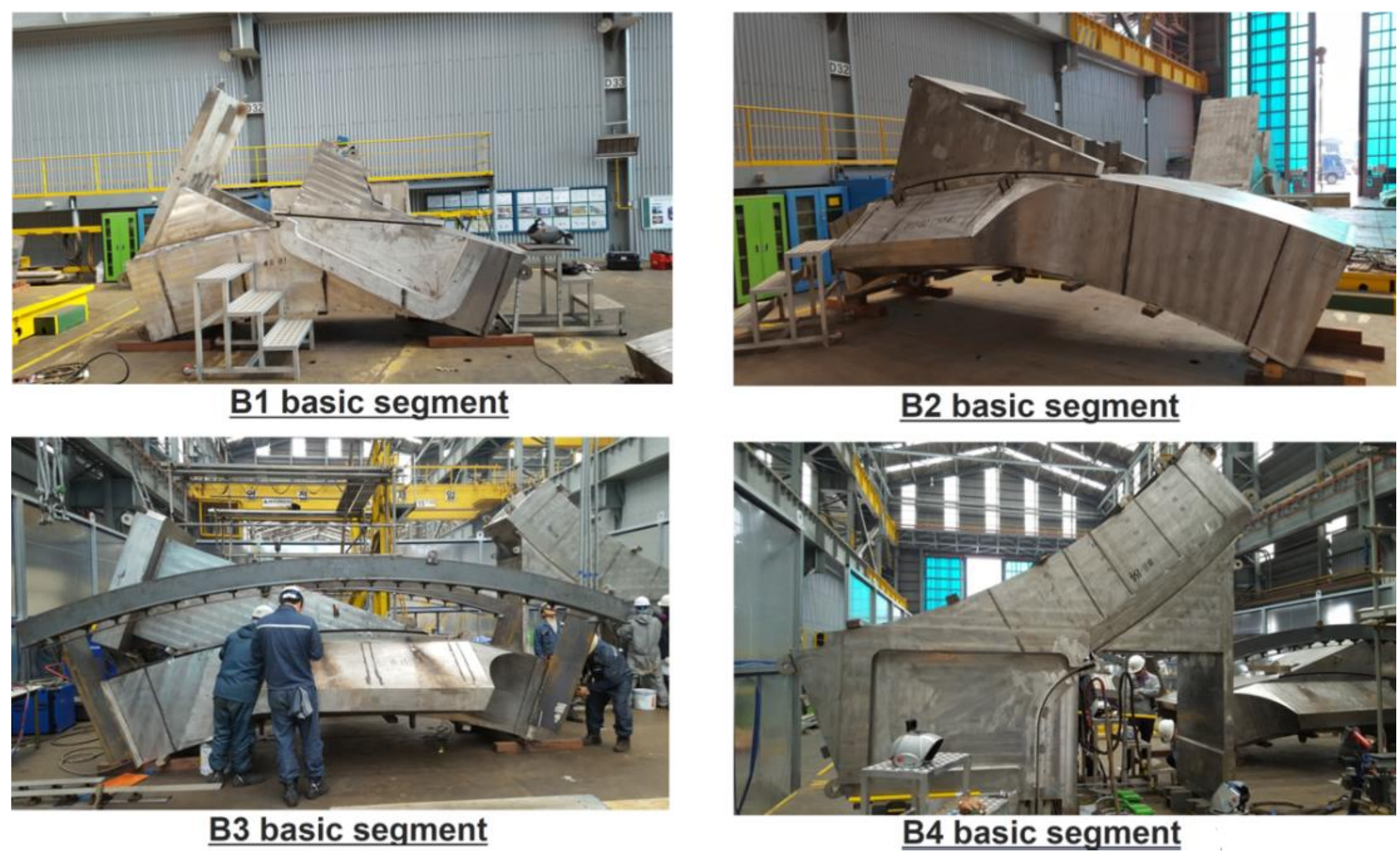

**Fig. 28:** Main TF Case Segments under Manufacture.

Because of the complex load conditions, efficient structural design is necessary to avoid large material volumes to satisfy the stress limits, and is difficult to achieve. Minor areas can dominate performance, with large volumes not near critical stresses.

## 5.2 Maintaining Tolerances and Overmetal

A crucial step, as a component is built up, is maintaining tolerances. Weld distortions are the main source and can be minimised but rarely eliminated. These tolerances are not just related to the final component but for alignment of weld interfaces for the next manufacturing step. Machining of course has its own tolerance limits (vibration and temperature control are critical) but generally weld distortions can only be recovered by leaving an extra thickness which can be (non-uniformly) machined to recover the correct dimensions.

This machining may occur in several steps, and the over-metal must match the expected distortions. For cost reasons, it is better to limit final machining to essential interface regions. As the component gets larger, so, generally, do the machine tools and the difficulty to achieve tight machining tolerances. As an example, the European Domestic Agency (EU-DA) for ITER did extensive weld distortion simulations and verified them by welding trials, before selecting the over-metal thicknesses shown in Fig. 29. Obviously, too much overmetal creates very large final machining costs and a fine judgement is required. A more relaxed First-of-Kind (FoK) manufacture (where the first components are used only as spares) allows for a refinement. All the ITER TF coil structures went through a final machining, in Japan and Europe, as illustrated in Fig. 30. With more time (and a full FoK) it is likely that this final machining could have been avoided.

The final step of the manufacturing is assembly into the FPP magnet system. At this point, the final magnet tolerances require assembly gaps between components, The problem with these tolerance gaps is what happens when they are closed. It is very difficult to shim large surfaces effectively, and the magnitude of the forces between individual magnets (Section 4) means that it is difficult to limit the area. ITER structural assessments included detailed modelling of gap closures, and the final assembly includes multiple matched fit keys and pins.

As an example, for the final machining of the TF coil F4E defined the extra material as

10mm Extra material on:

- PF interfaces (PF1-2-5-6)
- IIS grooves
- PCCF holes
- Straight leg
- IOIS surface and hole's radius
- OIS surface and hole's radius
- BU vertical tooling assembly surface
- Gravity support surface

3mm Extra material on:

- AU-BU Butt weld interfaces

No extra material for:

- CC interfaces
- Assembly holes interfaces
- ILIS holes

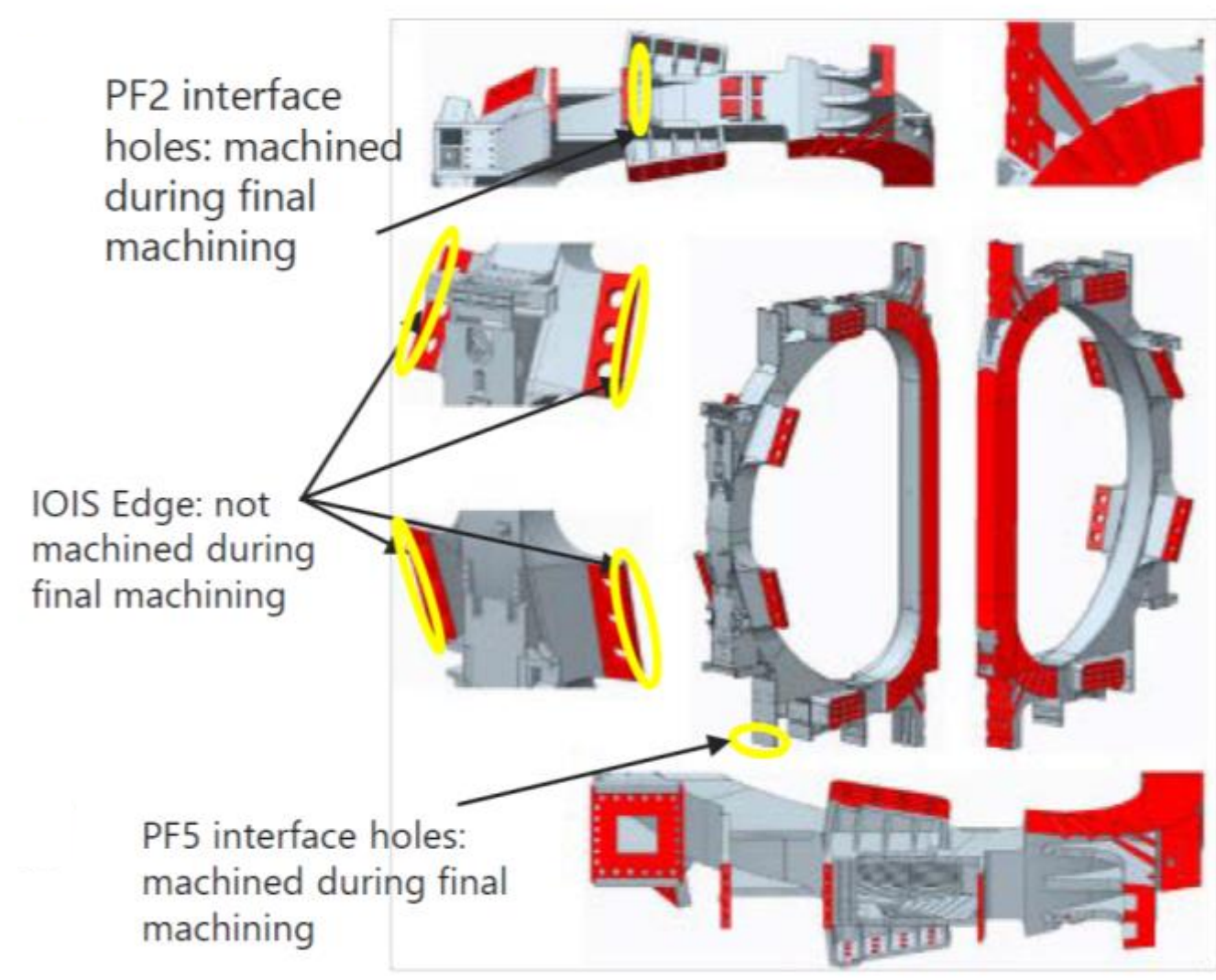

TF Coil: Red interface areas for final machining 24

**Fig. 29:** Over-metal regions in the TF coil.

Although not originally intended, both JA-DA and EU-DA used large gantry machines for a final machining of the coils, as shown in Fig. 30.

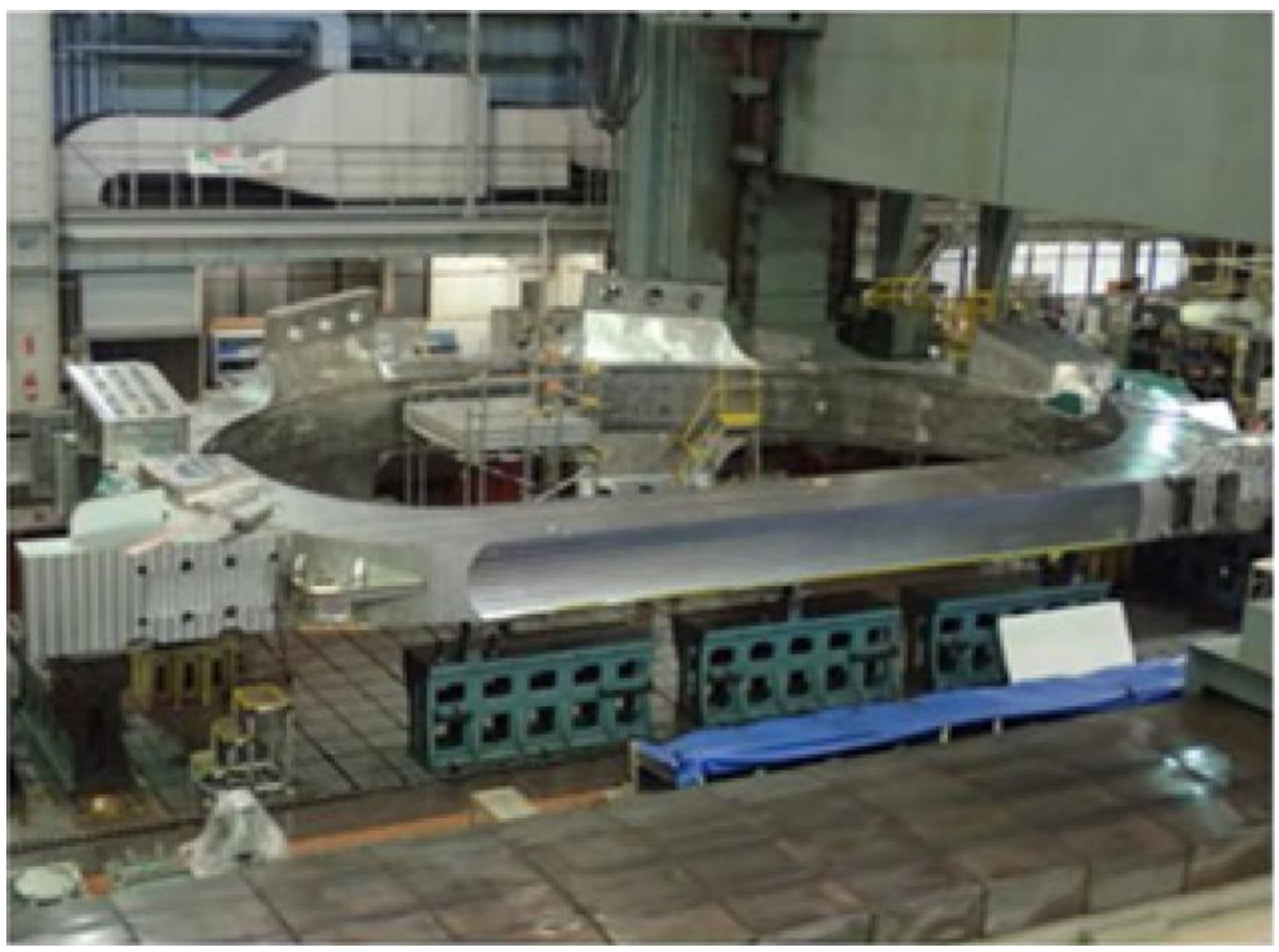

**Fig. 30:** TF coil on the Super Miller at MHI.

The magnet structures for ITER turned out to be one of the most difficult procurements of the magnets (more than the conductors), from a combination of sophisticated technical requirements not fully anticipated by the Domestic Agencies involved (nor by the IO) and the ITER procurement sharing, which multiplied interfaces and routes, forcing multiple process qualifications all with their own individual problems.

## 6 Conclusions

Basic structural materials (essentially austenitic steels) suitable for cryogenic magnets have been available in their present forms for about 50 years but have in the last 20 years seen very large improvements in their industrial fabricability (and hence cost and performance). This comes from a very wide user base in other technical areas that is funding tooling improvements in refining, forging and welding. This can, with care, be exploited in new FPP to improve on ITER selections. With care means being at the front of the technology, but not ahead of it, and ensuring that an adequate supply chain exists for the technology level selected. Some of ITER's problems were due to being ahead of standard industrial practice.

At a basic, almost pre-concept level, the selection of structural forms (i.e. the way to include structural material) subsequently impacts the entire FPP design (also non-magnet areas). It is a balance between relatively low cost bulk material that is used in a structurally inefficient way (and so has more weight and volume), and more targeted distributions of material that require more forming and better tolerances, so less weight but a higher unit cost. In ITER, the choice was substantially for relatively low cost bulk material, although with one or two areas of innovation (that could be developed).

There are two routes to achieve more efficient structural design:

i. The present link between the magnet mechanical design criteria and pressure vessel codes is unhelpful and forces excessive conservatism on the analysis and the criteria. There is an opportunity (a narrow one) to develop the ITER magnet mechanical criteria into a modernised set based on modelling of imperfections and the way they create failures (i.e much closer to what a digital twin should be providing, but rarely does). Thus, the structural analysis will include tolerances, gaps, sliding, bonded insulation, residual stresses as well as defects, fatigue and plasticity/work hardening. Failure is no longer based on empiricism around bulk conditions but detailed material modelling.

ii. The tendency of magnet designers in the past (including those on ITER) to react large forces with large structures should be reviewed. Magnets are inherently composite structures involving distributed structural materials as well, and it may prove more efficient to integrate the structural material more directly into the coil winding and eliminate huge forged & machined cases. Essentially, gluing is much cheaper than welding.

A number of innovative magnet technologies are appearing, in particular REBCO HTS materials (allowing of course higher fields but more importantly, higher temperatures (40K at 12T is possible, instead of 7K with Nb3Sn) and removing the $Nb_3Sn$ reaction heat treatment. Then demountable coils are making an appearance in tokamak and stellarator power plant designs. These bring a vast amount of flexibility to the magnet concept (and would have led to a radically different design for ITER if they had been available) and give many novel possibilities for structural support, Fig. 31. Quench protection is developing to bring much lower voltages to large high energy magnet designs, it being recognised that high voltage, cryogenic vacuum and stressed insulation create multiple new failure routes. There is an opportunity to move forward on innovative magnet design criteria for these new magnets.

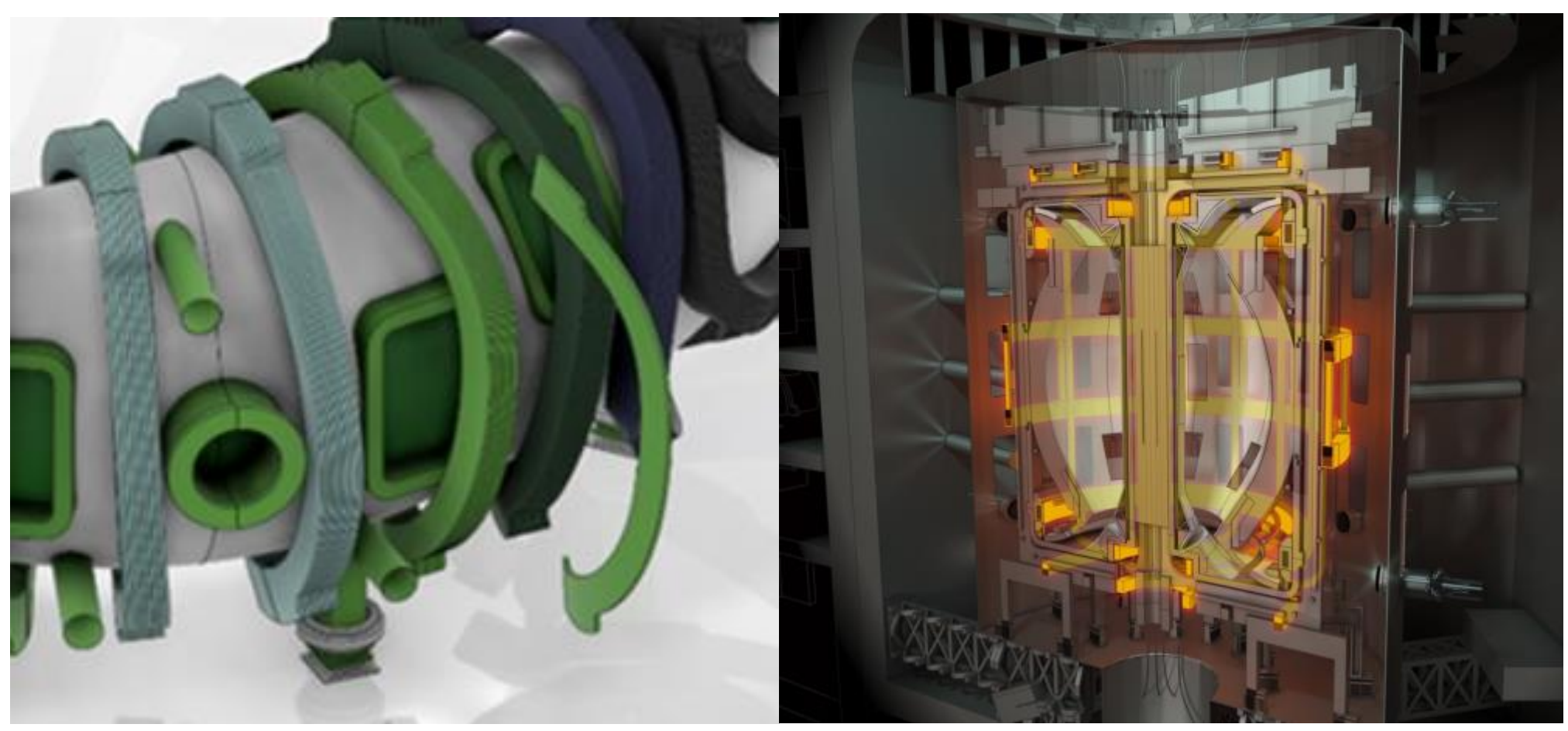

**Fig. 31:** Left, Gauss Fusion Stellarator Power Plant with Jointed Coils, Right STEP UKAEA Spherical tokamak demonstration power plant, with Jointed Coils [37, 38].

## Acknowledgement

Much of the detail (and the lessons learned) come from the ITER project and the contributions of my colleagues in the magnet division over the years is gratefully acknowledged.